\documentclass[12pt]{emulateapj}

\def \HST{{\emph{HST}}}
\def \Spitzer{{\emph{Spitzer}}}
\def \Herschel{{\emph{Herschel}}}

\def \tf {{24 $\mu$m}}
\def \um {{$\mu$m}}
\def \boot {{Bo\"otes}}

\def \lsun {{$L_{\odot}$}}
\def \msun {{$M_{\odot}$}}
\def \lir {{$L_{IR}$}}

\def \yr {{yr$^{-1}$}}

\begin{document}
\slugcomment{3/3/12}

\title{The Spectral Energy Distributions and Infrared Luminosities of $z\approx 2$ Dust Obscured Galaxies from \Herschel\ 
and \Spitzer
}
 
\author{J. Melbourne \altaffilmark{1}, B. T. Soifer \altaffilmark{1,2}, Vandana Desai \altaffilmark{2}, Alexandra Pope \altaffilmark{3}, Lee Armus \altaffilmark{2}, Arjun Dey \altaffilmark{4}, R. S. Bussmann \altaffilmark{5}, B. T. Jannuzi \altaffilmark{4}, Stacey Alberts\altaffilmark{3}}

\altaffiltext{1}{Caltech Optical Observatories, Division of Physics, Mathematics and Astronomy, Mail Stop 320-47, California Institute of Technology, Pasadena, CA 91125, jmel@caltech.edu,  bts@submm.caltech.edu}

\altaffiltext{2}{Spitzer Science Center, Mail Stop 314-6, California Institute of Technology, Pasadena, CA 91125, bts@ipac.caltech.edu,  lee@ipac.caltech.edu, vandesai@gmail.com}

\altaffiltext{3}{University of Massachusetts, Astronomy Department, Amherst MA, pope@astro.umass.edu}

\altaffiltext{4}{National Optical Astronomy Observatory, P.O. Box 26732, Tucson, AZ 85726-6732, dey@noao.edu, jannuzi@noao.edu}

\altaffiltext{5}{Harvard/Smithsonian Center for Astrophysics, Cambridge MA, rbussmann@cfa.harvard.edu}

\begin{abstract}
Dust-obscured galaxies (DOGs) are a subset of  high-redshift ($z\approx 2$) optically-faint ultra-luminous infrared galaxies (ULIRGs, e.g. $L_{IR} >  10^{12}$ \lsun). We present new far-infrared photometry, at 250, 350, and 500 \um\ (observed-frame), from the {\it Herschel} Space Telescope for a large sample of 113 DOGs with spectroscopically measured redshifts. Approximately 60\% of the sample are detected in the far-IR.  The \Herschel\ photometry allows the first robust determinations of the total infrared luminosities of a large sample of DOGs, confirming their high IR luminosities, which range from $10^{11.6}$ \lsun $< L_{IR} (8-1000 $\um$) <10^{13.6}$  \lsun. 90\% of the \Herschel\ detected DOGs in this sample are ULIRGs and 30\% have $L_{IR}  > 10^{13}$ \lsun. The rest-frame near-IR ($1 - 3$ \um) SEDs of the \Herschel\ detected DOGs are predictors of their SEDs at longer wavelengths.  DOGs with ``power-law'' SEDs in the rest-frame near-IR  show observed-frame 250/24 \um\ flux density ratios similar to the QSO-like local ULIRG, Mrk~231. DOGs with a stellar ``bump'' in their rest-frame near-IR  show observed-frame 250/24 \um\ flux density ratios similar to local star-bursting ULIRGs like NGC~6240.  None show 250/24 \um\ flux density ratios similar to extreme local ULIRG, Arp~220; though three show 350/24 \um\ flux density ratios similar to Arp~220.  For the \Herschel\ detected DOGs, accurate estimates  (within $\sim25$\%) of total IR luminosity  can be predicted from their rest-frame mid-IR data alone (e.g. from \Spitzer\ observed-frame 24 \um\ luminosities).  \Herschel\ detected DOGs tend to have a high ratio of infrared luminosity to rest-frame 8$\mu$m luminosity (the $IR8= L_{IR}(8-1000 \mu m)/\nu L_{\nu}(8 \mu m)$ parameter of Elbaz et al. 2011). Instead of lying on the $z=1-2$ ``infrared main-sequence" of star forming galaxies (like typical LIRGs and ULIRGs at those epochs) the DOGs, especially large fractions of the bump sources, tend to lie in the starburst sequence.  While, \Herschel\ detected DOGs are similar to scaled up versions of local ULIRGs in terms of 250/24 \um\ flux density ratio, and $IR8$, they tend to have cooler far-IR dust temperatures ($20-40$ K for DOGs vs. $40-50$ K for local ULIRGs) as measured by the rest-frame 80/115 \um\ flux density ratios (e.g., observed-frame 250/350 \um\ ratios at $z=2$). DOGs that are {\emph{not}} detected by \Herschel\ appear to have lower observed-frame 250/24 \um\ ratios than the detected sample, either because of warmer dust temperatures, lower IR luminosities, or both.      
\end{abstract}


\section{Introduction}


A very simple optical to mid-infrared (mid-IR) color selection of $R-[24] > 14$ (Vega mags, i.e., $F_{\nu}$(\tf)$ / F_{\nu} (R) \ga 1000$) yields a sample of optically-faint ultra-luminous infrared galaxies (ULIRGs, $L_{IR} > 10^{12}$ \lsun) at $z\sim2$ \citep[e.g.,][]{Houck05,Yan07, Dey08, Fiore08, Lonsdale09, Donley10}.  Galaxies selected this way have been termed dust obscured galaxies (DOGs), and they are among the most luminous galaxies at their redshift. Large 24 \um\ flux densities imply dust heating either by significant star formation, AGN activity, or both.  However, until recently, there have been few actual constraints on the total infrared (IR) luminosities, \lir (8-1000 \um), of the DOGs, because of a lack of deep observations across the far-infrared dust peak (e.g. rest-frame $60 - 200$ \um).  In this paper, we use \Herschel\ SPIRE \citep{Griffin10} observations at 250, 350 and 500 \um\ from the \Herschel\ Multi-tiered Extragalactic Survey \citep[HerMES;][]{Oliver10} to trace the far-infrared (far-IR) spectral energy distributions (SEDs) and infrared (IR) luminosities of a large sample (113) of DOGs with measured spectroscopic redshifts. 

We have identified over 2600 DOGs \citep{Dey08} selected from a \Spitzer\ 24 \um\ imaging survey  of  $\sim9$ square degrees in \boot\ \citep{leFloch_inprep}.  The \Spitzer\ program reached a flux limit of 0.3 mJy at 24 \um, and overlapped the deep optical imaging program, from the NOAO Deep Wide Field Survey \citep[NDWFS][]{JannuziDey99}.  Redshifts for over 100 DOGs have been obtained from spectroscopy campaigns using the Keck 10m, Palomar 5m, and \Spitzer\ Space Telescope \citep{Houck05, Weedman06, Brand07, Desai07, Dey08, Melbourne11}.  The spectroscopic surveys show a surprisingly narrow redshift distribution for the DOGs, with a mean $z\simeq2.0\pm0.5$ for the sample. 

The rest-frame near-infrared (near-IR) spectral energy distributions of DOGs measured by the \Spitzer\ Infrared Array Camera \citep[IRAC,][]{Fazio04} yield two classes.  The fainter 24 \um\  sources (e.g. $< 0.8$ mJy) tend to show a rest-frame 1.6 \um\ ``bump'' in their spectral energy distributions indicative of the photospheres of late type stars.  Mid-IR spectroscopy of the ``bump'' DOGs from the \Spitzer\ Infrared Spectrograph \citep[IRS,][]{Houck04} show strong polycyclic aromatic hydrocarbon (PAH) emission, which are typically found in galaxies with ongoing star-formation \citep{Yan07, Desai07, Huang09}.  The brighter DOGs tend to show a rising power-law SED in the near-IR-to-mid-IR bands.  \Spitzer\ IRS spectra of these ``power-law'' DOGs generally lack PAH emission, and instead show a rising continuum, an indicator of warm dust.  This lack of PAH emission and significant warm dust is usually taken as a sign of AGN activity \citep{Houck05, Weedman06, Yan07}. Many of these power-law DOGs also show deep silicate absorption in their IRS spectra suggesting high levels of dust obscuration. Rest-frame optical spectroscopy of the power-law DOGs reveals further evidence for AGN activity via broad $H\alpha$ emission lines \citep{Brand07,Melbourne11}.  AGN activity has also been inferred from the X-ray hardness ratio for both stacked \citep{Fiore08} and individual sources  \citep{Melbourne11}. 

 \HST\ and Keck Adaptive Optics Images have revealed the rest-frame UV--optical morphologies of DOGs, which range from compact (and point-like), especially for the more luminous power-law sources, to diffuse and/or more disk-like for the less luminous bump DOGs \citep{Melbourne08b, Bussmann09, Melbourne09, Donley10,Bussmann11}.  Some DOGs show clear signs of recent merging \citep{Dasyra08, Melbourne09, Donley10, Bussmann11}, but for many the evidence for an ongoing merger is marginal at best.   
 
 The number densities and clustering strength of DOGs are similar to sub-mm galaxies (SMGs) and high-z QSOs suggesting the possibility of an evolutionary connection \citep{Chapman05, Brodwin08, Chapman09}. In fact, there is some overlap between DOG and SMG selections ($\sim30$\%) especially at fainter 24 \um\ flux densities \citep{Pope08}. These results suggest that the DOGs likely occupy relatively massive halos and may evolve into today's $3-7$ $L^*$ galaxies \citep{Brodwin08}.    

These observational results have informed theoretical models of the DOGs.  To achieve the high mid-IR luminosities, modelers often invoke galaxy gas-rich major mergers \citep[e.g.][]{Mihos96}. In such models, a merging system can evolve through several periods of very high mid-IR luminosity that result in a  DOG classification \citep{Narayanan10}.  During final coalescence, star formation rapidly increases and the system can be simultaneously classified as a bump DOG and/or an SMG.  Eventually, black-hole growth starts to pick up, and star formation begins to slow. During this phase the galaxy may become a power-law DOG, before eventually settling into a massive quiescent galaxy.  While this theoretical picture may explain these classes of extreme $z=2$ galaxies, current observations cannot link these high-z galaxies in a causal chain, or even place the bulk of them in mergers.  However, it is a helpful framework for understanding the types of processes that can lead to these systems.  

While much is now known about the DOGs, a key missing piece of information has been a direct measurement of their total IR luminosities.  Unlike SMGs most of the luminous DOGs in our sample have been difficult to observe in the sub-mm \citep{Pope08, Bussmann09b}.  Thus their total IR luminosities have not been well constrained.  The lack of detections in the sub-mm bands suggests that their dust temperatures may be warmer than the typical SMGs \citep{Kovacs06, Coppin08, Sajina08, Younger09, Lonsdale09, Bussmann09b, Fiolet09}.  Likewise these galaxies have been difficult to detect in the longer \Spitzer\ bands  \citep[70 and 160 \um,][]{Tyler09}.  However, with the deep \Herschel\ SPIRE observations of the \boot\ field at 250, 350 and 500 \um\ \citep[from the HerMES team,][]{Oliver10}, strong constraints can finally be placed on the far-IR SEDs and IR luminosities of a large sample of DOGs.

In this paper, we investigate the optical through far-IR SEDs of 113 DOGs with known spectroscopic redshifts that lie in the \Herschel\ fields.  We measure the far-IR flux densities from the \Herschel\ SPIRE observations, and compare to the SEDs of three local ULIRGs that range from AGN dominated to star formation dominated.  We use the SPIRE observations  to constrain the total IR luminosities (8 -1000 \um) and far-IR temperatures of the DOGs, and compare with other $z=1-2$ ULIRG and AGN samples.  

Section 2 describes the sample, the \Herschel\ observations, and far-IR photometry.  Section 3 presents the observed SEDs, SED classifications, IR luminosities, and far-IR dust temperatures.  Section 4 discusses the results in the context of other high-$z$ galaxy samples. Section 5 summarizes our conclusions.  Throughout we assume the canonical $\Lambda$ Cold Dark Matter Universe with $\Omega_M=0.3$, $\Omega_{\Lambda}=0.7$, and $H_0=70$ km s$^{-1}$ Mpc$^{-1}$.

\section{Sample Selection and Observations}
The sample of DOGs is taken from the \boot\ field of the NDWFS \citep{JannuziDey99}.  This field, roughly 9 square degrees in area, was observed with \Spitzer\ Multiband Imaging Photometer for Spitzer \citep[MIPS,][]{Rieke04} at 24 \um, reaching an 80\% completeness depth of 0.3 mJy \citep{leFloch_inprep}. The field also has deep optical imaging in the $B_W$, $R$, $I$, and $K$ bands to depths of 27.1,	26.1,	25.4, and 19.0 mag (Vega) respectively.  Moderately deep \Spitzer\ IRAC imaging at 3.5, 4.6, 5.8, 8.0 \um\ was obtained for the entire field \citep{Eisenhardt04} and augmented by the \Spitzer\ Deep Wide-Field Survey \citep[SDWFS;][]{Ashby09}.   

\begin{figure*}
\centering
\includegraphics[scale=0.8]{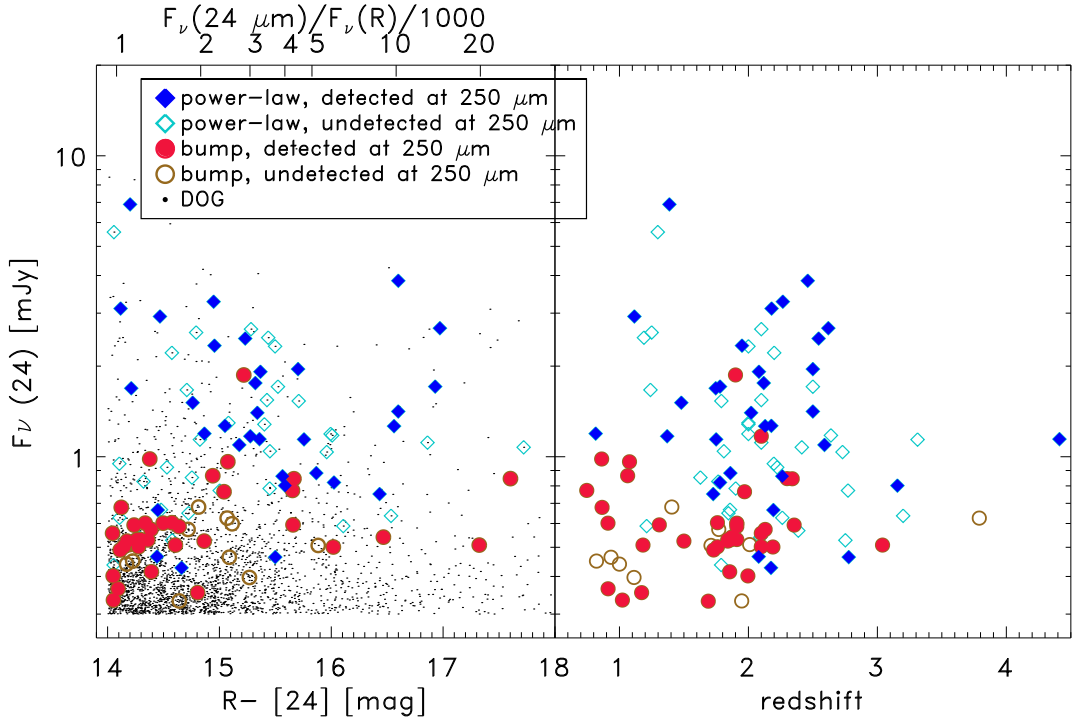}
\caption{\label{fig:sample} Left: $F_\nu (24)$ [mJy] vs. $R-[24]$ [Vega mag] for the complete sample of DOGs in \boot\ (points) and those with spectroscopic redshifts (symbols) divided into rest-frame near-IR classifications of bump DOGs (circles) and power-law DOGs (diamonds). Right: $F_\nu (24)$ [mJy] vs. redshift.  DOGs are selected to have $R-[24] > 14$ [mags] (i.e., $F_{\nu}$(\tf)$ / F_{\nu} (R) \ga 1000$). The spectroscopic samples roughly span the full range of $R-[24]$ color for the larger sample, although the bump DOGs  tend to be drawn from the bluer end of the distribution. The spectroscopic samples tend to be drawn from the brighter end of the sample, especially for the power-law DOGs.  There are no obvious trends in 24 \um\ flux density with redshift. Likewise there are not obvious trends for \Herschel\ detected (filled symbols) vs. \Herschel\ non-detected (open symbols) sources.
}
\end{figure*}

The large survey area was key for identifying statistically significant samples of rare yet luminous sources. Of the $\sim2600$ DOGs in \boot\, spectroscopic redshifts were obtained for 117 galaxies \citep{Houck05, Weedman06, Brand07, Desai08, Dey08,Melbourne11}.  In all cases where spectra yielded redshifts, the DOGs have been found to lie in a relatively tight redshift range of $<z> = 2.0$, $\sigma_z=0.5$.  

The \boot\ field has been observed at longer wavelengths with \Herschel\ SPIRE at 250, 350, and 500 \um\ as part of the HerMES collaboration \citep{Oliver10,Brisbin10, Rigopoulou10, Seymour11}. This paper presents results from the far-IR \Herschel\ observations of 113 DOGs with spectroscopic redshifts.  This is not a statistically complete sample of DOGs, but it is representative of the more luminous DOGs observed in the \boot\ field as shown in Figures \ref{fig:sample} and \ref{fig:IRAC}.  

The sample used in this paper includes 86 of the 90 DOGs studied in \citet[][which placed constraints on the stellar masses of the DOGs]{Bussmann11b}.  Four of the Bussmann et al. DOGs lie off of the SPIRE mosaics and so are not included in this study. We also include 27 additional DOGs with redshifts below the Bussmann et al. redshift limit of $z=1.4$.  Table \ref{tab:sample} gives the R.A. and Dec., redshifts, the 24 \um\ flux densities, and the $R-[24]$ colors of the sample of 113 galaxies. Table \ref{tab:bluephot} gives the optical through mid-IR flux densities of the sample.  

\subsection{Rest-frame Near-IR SED Classification}
As was described in the introduction, DOGs show two types of rest-frame near-IR SEDs, ``power-law'' sources with a rising SED across the Spitzer IRAC bands, and``bump'' sources with a peak or break in their SED across the IRAC bands.  This bump has been associated with the photospheres of late-type stars, and appears at rest-frame 1.6 \um.  Distinguishing between bump vs. power-law samples is complicated by the bump shifting in the observed \Spitzer\ bands for objects at different redshift.  

\begin{figure}
\includegraphics[scale=0.5]{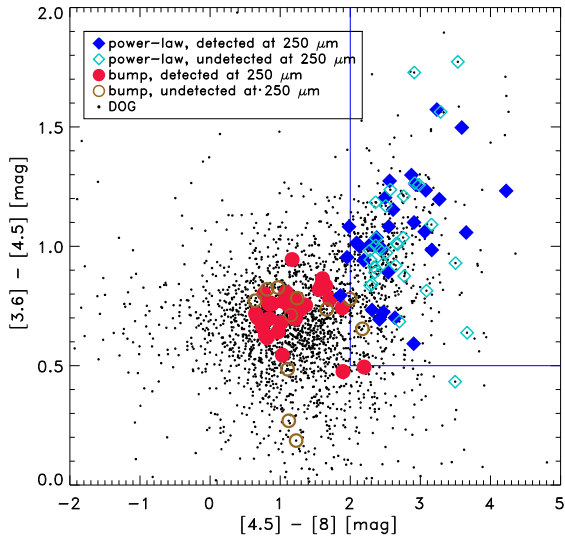}
\caption{\label{fig:IRAC} The \Spitzer\ IRAC infrared color-color plot of all the DOGs in \boot\ (points) and the spectroscopic samples (symbols).  DOGs with a rising ``power-law'' SED in the IRAC bands (diamonds) segregate from the DOGs with a ``bump'' in their SED at rest-frame 1.6 \um\   (circles). The power-law sources tend to be red in both the [3.6] - [4.5] and [4.5] - [8.0] colors.  Detection in \Herschel\ does not appear to be driven by IRAC colors (filled vs. open symbols). Although, the power-law DOGs in this sample are less likely to be detected by \Herschel\ than the bump dogs.   }
\end{figure}

The SEDs were visually classified into bump vs. power-law, based on the rest-frame $1-8$ \um\ SED.  Sources with a clear 1.6 \um\ peak in their SED were selected as bump sources.  The two samples are well segregated, in IRAC color-color space as shown in Figure \ref{fig:IRAC}.   The power-law sources are red in both [3.6] - [4.5] color and [4.5] - [8.0] color.  In contrast, the bump sources tend to be fairly blue in [4.5] - [8.0] color. The near-IR SED classifications are given in Table \ref{tab:photometry}.  58\% of the spectroscopic sample are power-law sources and 42\% are bump sources.  Again these fractions are only representative of this sample and not the larger DOG population which appear to favor bump sources especially at the lower 24 \um\ flux density levels (e.g., Figure \ref{fig:sample}).

\citet{Bussmann11b} also provides a rest-frame near-IR classification based on the IRAC photometry.  These previous efforts used linear fits to the IRAC data to classify the DOGs and were designed to statistically separate out the two classes.  Even though these previous classifications did not consider the redshift dependence of the position of the stellar bump they still agree with the new visual classifications for 89\% of the sample.  For the 11\% where the two classifications disagree, we have chosen to use the visual classification, because it accounts for differences in redshift.      

\subsection{\Herschel\ Far-IR Observations}
As part of the \Herschel\ GTO time, the \boot\ field was observed with the SPIRE far-IR imager by the HerMES team (P.I. Oliver).  The central 2 square degrees were observed to a depth of $\sim80$s  in all three SPIRE filters (250, 350, and 500 \um).  An additional annulus, with an outer diameter  of $\sim3$ degrees surrounding this central field, was imaged to a shallower depth of $\sim30$s, again in all three SPIRE filters.  These images were processed through the \Herschel\ Level 1 data reduction pipeline and  made publicly available.  The pipeline reduced images were used here to measure the far-IR flux densities of the DOGs in \boot.   

We combined the SPIRE observations into 250, 350, and 500 \um\ mosaiced images using the SWarp package \citep{Bertin02}. Image alignment was set by the header world coordinate system assigned to the images from the data reduction pipeline. These were adequate to align the images to sub-pixel precision, without significant loss of resolution. The same SWarp parameters were used to mosaic the instrument noise images.  The noise image mosaics were used to determine the formal photometric uncertainty of each measured galaxy as described below.

 

\subsection{Photometry}

While this paper is only concerned with the SPIRE photometry for the 113 DOGs in this sample,  we chose to generate a complete catalogue of point-sources in the SPIRE mosaics.  This approach allowed for better characterization of the photometric uncertainties and detection limits, as well as the alignment between the \Herschel\ and \Spitzer\ images.  

Photometry of the \Herschel\ mosaic images was carried out with a two step process.  First, the DAOphot \citep{Stetson87} FIND routine was used to identify sources in each mosaic image. FIND selects point-like sources in the signal maps.  The detection threshold was set low (e.g., 2 $\sigma$ above the noise level) to allow for the largest possible number of matches between the far-IR \Herschel\ data and the mid-IR \Spitzer\ data.  Second, the IDL 2D gaussian fitting routine, MPFit2DPeak (written by Craig Markwardt), was used to determine the flux density of each source. Throughout the mosaic process, the images retained the original flux density units of Jy/ beam. Thus the flux density of a point source in Jy is given by the peak value of a Gaussian fit to the source.  MPFit2DPeak returns the peak pixel value and the formal uncertainty for each measurement based on the instrument noise image and flux density level of the peak.  

These methods were used to measure 15748, 9118, and 5281 sources with flux densities greater than 20 mJy in the 250, 350, and 500 \um\ mosaics respectively.  Figure \ref{fig:detect} (upper panel) shows the flux density distribution of these detections.  

\subsection{Artificial Source Tests}
The photometric accuracy (limited by flux boosting from source confusion) and precision (i.e. photometric noise) were determined by populating the images with artificial sources and recovering their fluxes. These tests accounted for the effects of source confusion and background variations from unresolved cirrus. Artificial sources were created by scaling a very luminous source of known flux from each input image.  Artificial sources were placed randomly across each image and their fluxes were measured at the input locations.  This approach was analogous to determining the photometry of the DOGs, because the location of each DOG is also known ahead of time from the \Spitzer\ data. This test was not designed to recover the completeness limit of the \Herschel\ images.  

\begin{figure}
\centering
\includegraphics[scale=0.5,trim=0 50 0 50]{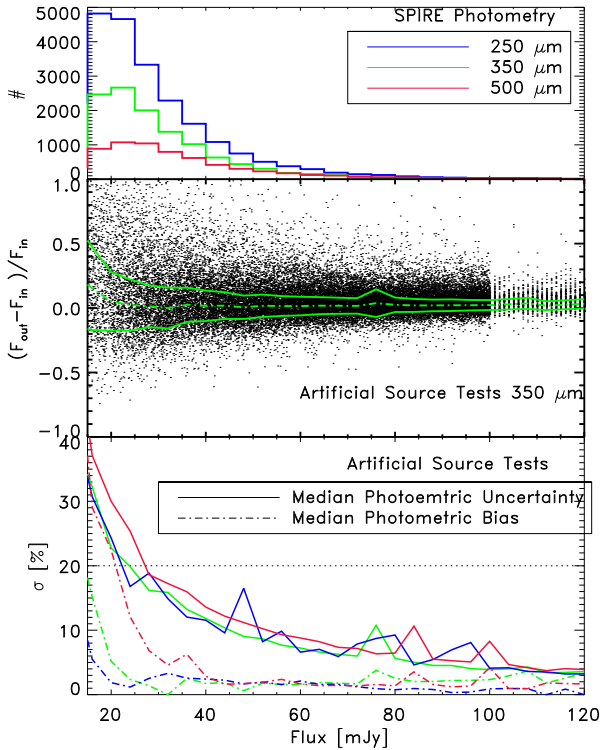}
\caption{\label{fig:detect} Top: Histograms of the sources counts in each of the \Herschel\ SPIRE filters as a function of the measured flux density in mJy. No corrections for flux boosting or confusion have been applied.  Middle: The fractional difference between input and output fluxes for artificial sources placed randomly  across the 350 \um\ SPIRE mosaic. The median (dot-dashed) and standard deviation (solid line) of the fractional differences are shown and represent the accuracy (which can be affected by flux boosting) and precision (photometric noise) of the photometry respectively. Bottom: The photometric accuracy (dot-dashed) and precision (solid lines) from artificial source tests on the 250 (blue), 350 (green) and 500 (red) \um\ images.    The photometry at 250 and 350 \um\ is good to within 20\% (dashed line) for galaxies with flux densities brighter than $\sim25$ mJy.  The 500 \um\ photometry is good to within 20\% for flux densities brighter than $\sim30$ mJy.  }
\end{figure}

Each mosaic was populated with 100,000 artificial sources, placed randomly, one at a time, so as to not increase the crowding. Input flux densities ranged from 400 mJy to 5 mJy. The flux density of each artificial source was measured with the same method as the real sources, including the same five arcsec positional threshold for matching the peak location.  Figure \ref{fig:detect} (middle panel) shows (measured flux - input flux) / (input flux) as a function of input flux.  The photometric precision is given by the standard deviation of the flux differences (solid lines in Figure \ref{fig:detect}, middle and bottom panels), whereas the photometric accuracy is given by the median of the flux differences  (dot-dashed lines in Figure \ref{fig:detect}, middle and bottom panels)

The bottom panel of Figure \ref{fig:detect} summarizes the results for the artificial source tests.   The photometric precision is better than $\sim20$\% at 25 mJy for the 250 and 350 \um\ images.  The 500 \um\ images show a 20\% uncertainty at 30 mJy.  In addition to the photometric noise, there is an increasing photometric bias (flux boosting) at fainter flux density levels, with the returned flux higher than the input flux.  This can be understood in the context of background confusion boosting the measured flux of the artificial source.  The photometric bias is smaller than 10\% at 20 mJy for the 250 and 350 \um\ images, and smaller than 10 \% at 25 mJy for the 500 \um\ image. No correction for this bias was applied to the final photometry.

These results summarize the typical uncertainties across the SPIRE images.  However, the true uncertainty of a given source will depend on the local confusion which might be better or worse than average.  The precision and accuracy of the DOG photometry could well be better than the numbers quoted above.  Not only are the DOG locations known, but the locations of other far-IR sources are known as well. If we run our artificial source tests in locations that exclude the locations of existing 24 \um\ sources (excluding locations within 2 pixels of known sources) then the photometry achieves a 20\% precision at roughly 20, 20, and 25 mJy (for the 250, 350, and 500 \um\ images respectively).  These levels represent a best case scenario, and we will take these as the canonical photometric upper-limits for DOGs that are undetected in the SPIRE images. 

\subsection{Catalogue Matching}
\begin{figure}
\centering
\includegraphics[scale=0.5]{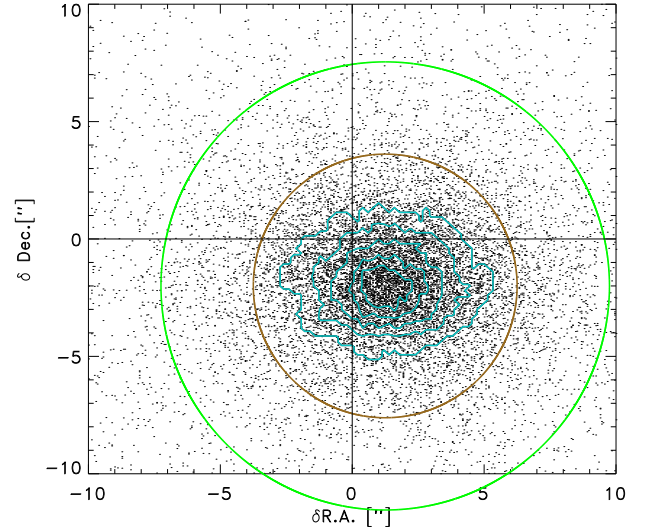}
\caption{\label{fig:match} The angular separation between 24 micron selected sources and 250 micron selected sources.  Cyan contours follow the density profile of the points and mark a positional offset between the two catalogues of  1.25$\arcsec$ and -2.0$\arcsec$ respectively in RA and Dec respectively.  A separation criteria of $5\arcsec$ (brown circle) recovers 80\% of the possible 24/250 \um\ matches that lie within the 250 PSF which has a half-width at half-maximum size of 8.5$\arcsec$ (green circle). 
}
\end{figure}

We match the full 250 \um\ catalogue to the full 24 \um\ catalogue \citep{leFloch_inprep} of the \boot\ field.  For each 24 micron source, the nearest 250 micron source was determined.  A plot of the difference in RA and Dec between the two catalogues (Figure \ref{fig:match}) reveals a linear spatial shift of 1.25$\arcsec$ and 2.0$\arcsec$  in RA and Dec respectively.   After applying these positional offsets to the \Herschel\ positions, matches were selected for objects with a separation of $<5 \arcsec$, roughly 1/3 of the 250 \um\ PSF size.   Of the 28391 24 micron sources (brighter than 0.3 mJy) in the \boot\ survey field, we find good SPIRE 250 \um\ matches (brighter than 20 mJy) for 6327 or 22\%.  

After matching the 24 \um\ sources to 250 \um\ counterparts, matches were made to sources in the longer wavelength data, based on the 250 \um\ positions.    
Approximately $\sim12$\% of the 24 \um\ sources have a counterpart at 350 \um, while $\sim6$\% have a 500 \um\ counterpart.

\begin{figure*}
\centering
\includegraphics[scale=0.65]{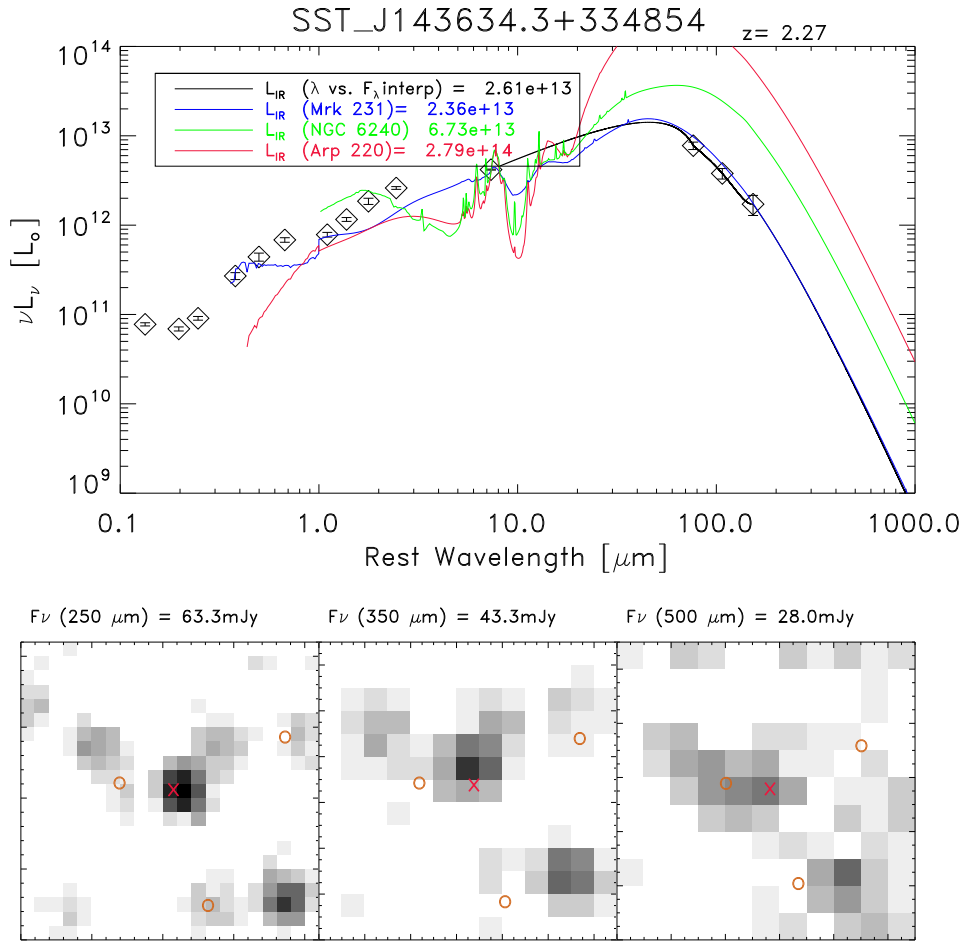}
\caption{\label{fig:SED} Top: $\nu L_\nu$ vs. wavelength for a power-law DOG (diamonds), and 3 local ULIRGs (lines).  Bottom: Postage stamps of the roughly $2\arcmin\times2\arcmin$ region around the DOG (marked with an x) from the 250 (left), 350 (middle), and 500 (right) \um\ SPIRE images. The postage stamp images also show the locations the neighboring 24 \um\ sources (o's).  While some sources suffer from blending most of the DOGs detected in SPIRE are relatively uncontaminated by neighbors.  SEDs of the DOGs are compared to templates of local ULIRGs, including Mrk~231 (AGN template in blue), NGC~6240 (a starburst template in green), and  Arp~220 (an extreme starburst in red).  The local templates are scaled to the 24 \um\ luminosities of the DOGs. Some of the DOGs (e.g., Figure \ref{fig:SED-a}), are better matched to the AGN dominated template, Mrk~231, while others are better matched to the starburst template, NGC~6240 (Figure \ref{fig:SED-b}).  None are well matched to the Arp~220 starburst, although some show similar 350 and  500 \um\ flux densities (Figure \ref{fig:SED-c}). Total \lir 's estimated from the scaled local templates are given in the legend.  When a template SED is well matched to the DOG data, the template derived \lir\ matches the \lir\ from a simple interpolation of the DOG SED in $\lambda$ vs. $F_\lambda$ space (black line). }
\end{figure*}

Finally, a visual check of the SPIRE images was made at the location of each of the 113 DOGs with redshifts to determine if those with measured far-IR flux densities show an actual source in the image, and that DOGs without a far-IR match do not show a significant source.  In all cases, a DAOphot detection resulted in a visually confirmed source (see Figure \ref{fig:SED}).  However, several DOGs that were undetected in the DAOphot catalogues did appear to contain a source at 250 \um.  Usually these were sources that were somewhat blended with a nearby neighbor causing the centroid of the final object to be offset from the 24 \um\ source at a larger separation than our match criteria of 5$\arcsec$.  For these cases, we fit the source by hand, forcing the centroid of the Gaussian to the position of the 24 \um\ detected DOG, and setting a background level to account for blended neighbor.   This ``by hand'' photometry was performed for 17 of the 113 DOGs in our sample.

Of the 113 DOGs in the sample, 68 (60\%) are detected at 250 \um, 56 (50\%) are detected at 350 \um, and 35 (31\%) are detected at 500 \um.  All of those DOGs detected at 350 and 500  \um\ are also detected at 250 \um.  The detection rate at 250 \um\ for 24 \um\ selected DOGs is significantly higher than for the 24 \um\ catalogue of \boot\ sources as a whole.

\section{Results}

\begin{figure*}[t]
\figurenum{5}
\centering
\includegraphics[scale=0.65]{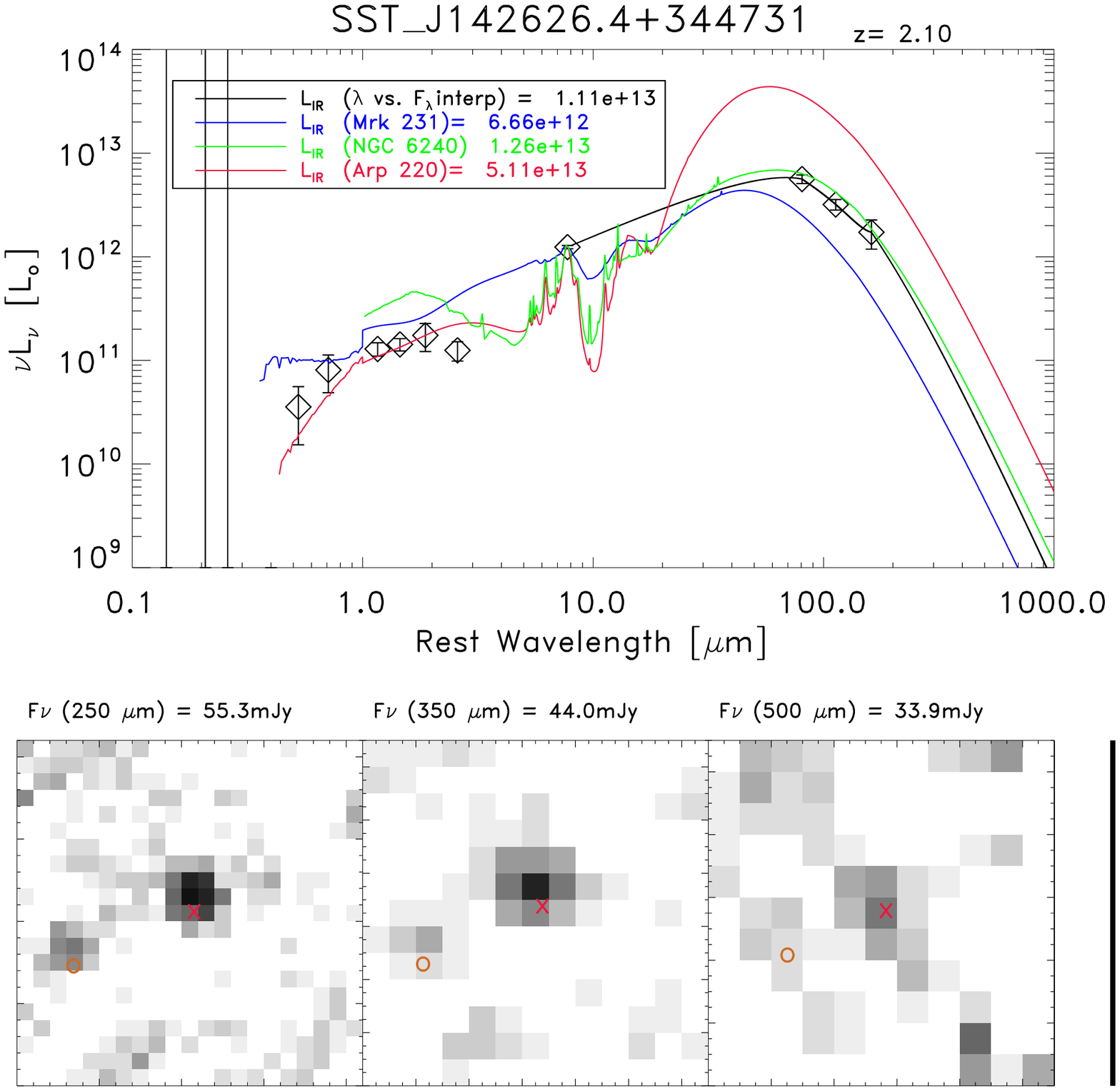}
\caption{Continued. SED and \Herschel\ SPIRE images for an IRAC classified bump DOG.  The far-IR SED of this galaxy is well matched to the starburst template NGC~6240.}
\end{figure*}

Figure \ref{fig:SED} shows the far-IR images, photometry measurements, and multi-wavelength SEDs for three of the  DOGs in our sample.  The locations of the DOGs in the Herschel images are marked with x's, while neighboring 24 micron sources are marked as o's.  The DOG SEDs are plotted in units of $\nu L_\nu$. In all cases where the DOGs are detected in SPIRE, the SEDs show a large far-IR peak associated with cold dust.  The \Herschel\ far-IR photometry and the \Spitzer\ 24 \um\ flux densities of the DOGs are given in Table \ref{tab:photometry}.

\subsection{Mid-to-Far-IR SED Classifications Based on the 250/24 \um\ Flux Density Ratio}

The mid-to-far-IR SEDs of the DOGs were classified by comparing them to scaled up versions of local ULIRGs (see Figure \ref{fig:SED}).  Mrk~231 is a Type-1 AGN-dominated ULIRG \citep{Sanders88}, although it also likely hosts some star formation \citep{Downes98, Davies04} which contributes to its far-IR flux at the 10-30\% level \citep{Armus07}. NGC~6240 is a starburst dominated ULIRG  \citep{Lutz03, Armus06}. It also hosts an AGN; however, the AGN contributes $<10$\% of the IR flux \citep{Max05, Armus06}.  Arp~220 is the nearest ULIRG and is also a  starburst. It possesses an extreme far-IR/mid-IR ratio, much larger than other local ULIRGs \citep{Armus07}.  Figure \ref{fig:SED} shows that the mid-to-far-IR SEDs of the DOGs span a range of shapes with some more like Mrk~231, and others resembling NGC~6240.  

\begin{figure*}[t]
\centering
\includegraphics[scale=0.8]{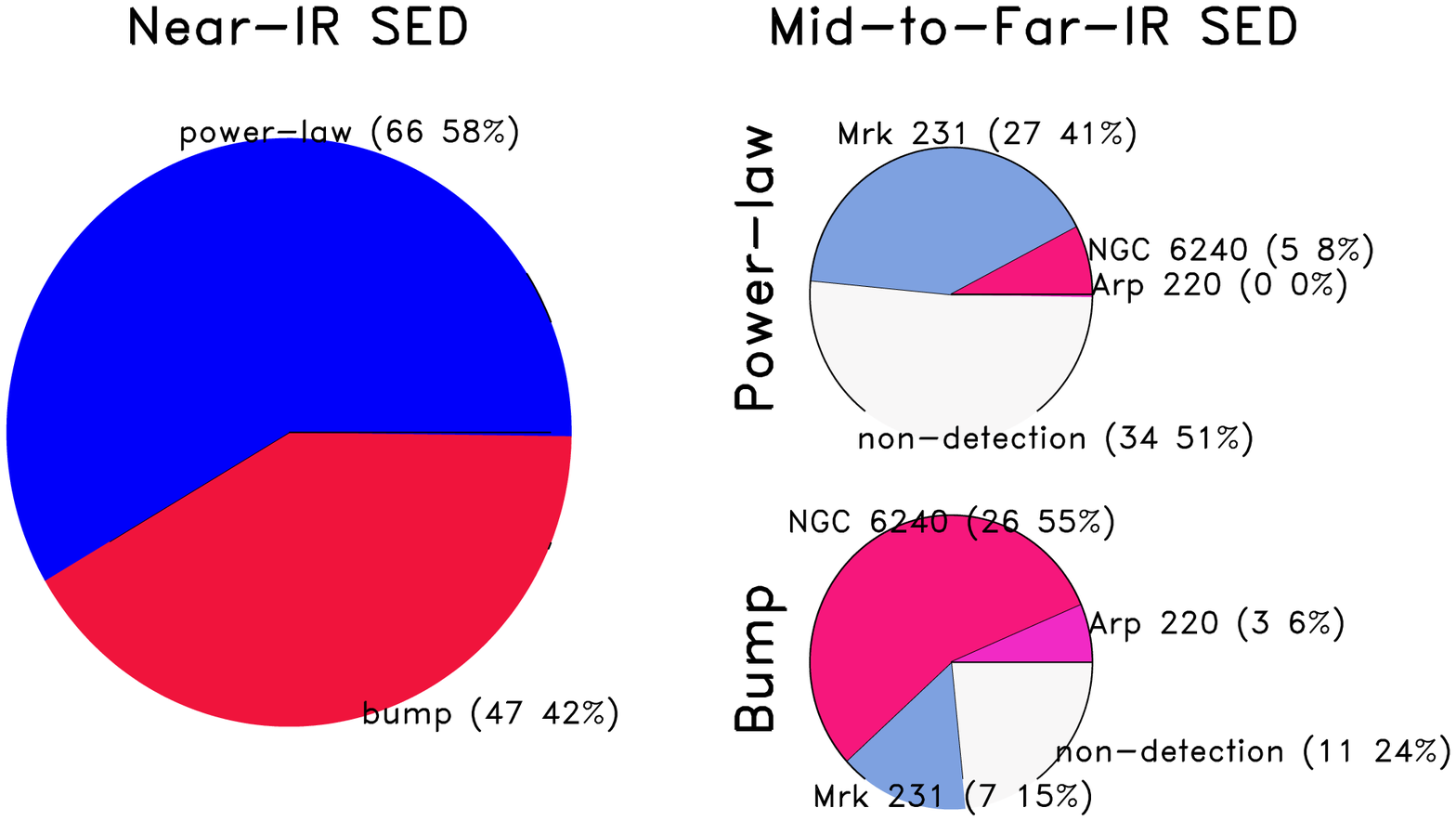}
\caption{\label{fig:FIRclass} Left: Fraction of the spectroscopic sample that are classified power-law vs. bump DOGs.  Right: classifications of their mid-to-far-IR SEDs.  Slightly more than half of the spectroscopic sample are power-law DOGs, with the remainder are classified as bump DOGs. However, we expect that a complete sample of DOGs will be dominated by lower-luminosity bump sources (e.g., Figure \ref{fig:IRAC}).  Over 50\% of the power-law DOGs are undetected at SPIRE wavelengths, while only $\sim$1/4 of the bump DOGs are undetected.  Of the power-law DOGs that are detected nearly all have mid-to-far-IR SEDs classified as AGN-like (Mrk~231).  Whereas, nearly all of the bump DOGs are classified as starburst like (NGC~6240 or Arp~220).}
\end{figure*}

As with the near-IR classifications, we first visually classify the mid-to-far-IR SEDs of the sample, based primarily on the observed 250/24 \um\ luminosity ratio.  Figure \ref{fig:FIRclass} shows the classification statistics for both bump and power-law DOGs.  Three results are immediately obvious from this Figure: (1) the power-law DOGs are less likely to be detected in the SPIRE bands than the bump DOGs, only 49\% of the power-law DOGs are detected, while 76\% of the bump DOGs are detected; (2) of the power-law DOGs that are detected, 84\% have AGN-like (Mrk~231) mid-to-far-IR SEDs; and (3) of the bump DOGs that are detected, 80\% have starburst-like (NGC~6240 or Arp~220) mid-to-far-IR SEDs.  The mid-IR-to-far-IR SED classifications for the full sample of SPIRE detected sources are given in Table \ref{tab:photometry}.

\begin{figure*}[t]
\figurenum{5}
\centering
\includegraphics[scale=0.65]{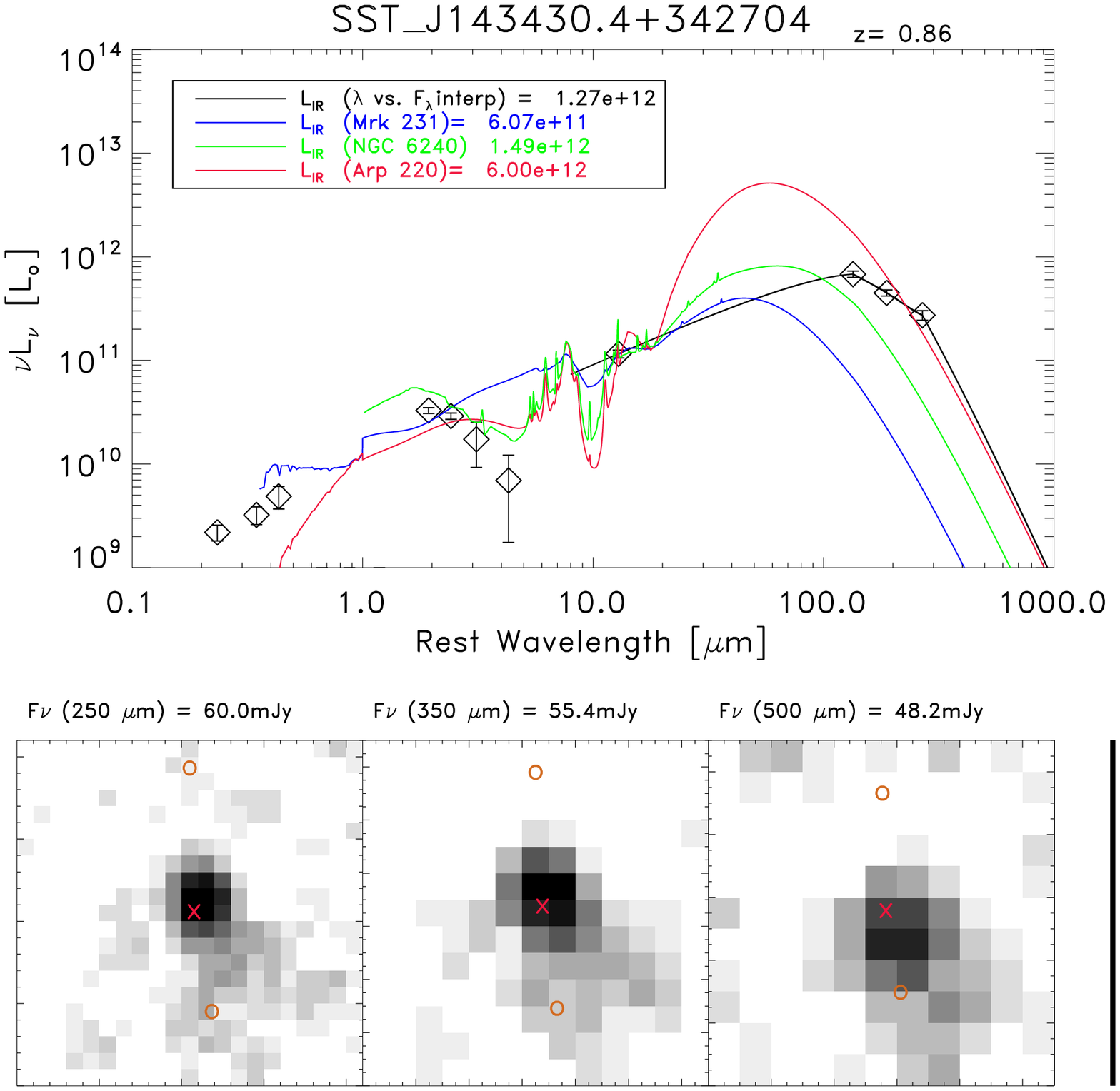}
\caption{Continued. SED and Herschel/SPIRE images for an IRAC-classified bump DOG.  This galaxy has an unusual FIR SED that somewhat resembles Arp~220, but with a lower flux density at 250 \um.}
\end{figure*}

The mid-to-far-IR SED classifications are being driven by the 250/24 \um\ flux density ratio.  To see this more easily, Figure \ref{fig:250to24} plots the 250/24 \um\ ratio for the DOGs as a function of redshift.  Over-plotted is this same ratio for the local templates redshifted to match the DOGs.  The power-law DOGs tend to have smaller 250/24 \um\ ratios than the bump sources, matching the redshifted 250/24 \um\ ratios of Mrk~231 (with significant scatter).  Similarly the bump DOGs match the redshifted 250/24 \um\ ratios of NGC~6240  (again with significant scatter), even across the $z=2$ redshift, where the 8 \um\ PAH features enter the 24 \um\ passband.
  
\subsection{Constraining the Total Infrared Luminosities, \lir ($8-1000$ \um)}

With the SPIRE far-IR observations we can, for the first time, observationally constrain the total infrared luminosities, \lir ($8-1000$ \um), of a large sample of DOGs.    However, even with the far-IR SPIRE observations, the SED is still only sampled at a few additional, though key, wavelengths.  Thus a measure of the total IR luminosity still requires some assumptions.  

We choose a simple approach for estimating IR luminosity. First we interpolate between the mid-IR and far-IR flux densities.  Then, for the long wavelength tail, we apply a black-body curve,  multiplied by $\nu^{1.5}$ to account for the dust emissivity \citep[see for instance][]{Draine03}.  We select a characteristic temperature for the far-IR tail of 40~K, although because the bulk of the luminosity is coming out at shorter wavelengths the total IR luminosity is relatively insensitive to the temperature used. A 25\% change in the far-IR temperature typically results in less than a 5\% change in the estimated luminosity. We interpolate the flux points in $F_{\lambda}$ vs. $\lambda$ space, which, as can be seen in Figure \ref{fig:SED} reproduces the shapes of the far-IR dust humps reasonably well. The resulting \lir\ measurements are tabulated in Table \ref{tab:photometry}. 

 As can be seen in Figure \ref{fig:SED}, when a DOG SED is well matched to a local template the \lir\ inferred for the DOG from the local template matches the \lir\ from this simple interpolation, to within better than 20\%.  Thus, while we could perform multi-component fits to our $2-4$ IR data points, the \lir\ measurements are unlikely to change significantly from this simple approach.
  
Figure \ref{fig:LIR} plots a histogram of the \lir\ measurements for both the power-law and bump DOGs detected in the SPIRE images.  Even though the power-law DOGs are less likely to be detected at 250 \um\ and have smaller 250/24 \um\ ratios, they tend to have higher luminosities than the bump sources.  While the bump sources typically have ULIRG luminosities of  \lir\ $=10^{12} -10^{13}$ \lsun, the power-law DOGs show a large fraction with \lir\ $> 10^{13}$ \lsun.   A K-S test reveals that the two distributions are extremely unlikely ($<1$\%) to be drawn from the same parent distribution. However, this is driven almost exclusively by the lack of lower luminosity power-law DOGs, which is most likely a selection bias.   Explanations for these results will be discussed in Section 4.

\subsection{Constraining the Far-IR Dust Temperature from the 250/350 \um\ Flux Density Ratios}


Most local ULIRGs cannot be fit by a single dust temperature \citep{Marshall07}, but rather contain both warm and cold components.  Because the DOGs are selected to be luminous at 24 \um\ (e.g., rest-frame 8 \um\ at $z=2$), they likely host significant amounts of warm and hot dust that will not be probed by the SPIRE observations.  However, the SPIRE measurements provide a characteristic temperature for the far-IR emission in the DOGs, which can be compared to the temperatures of other samples measured in the same way.

\begin{figure}
\centering
\includegraphics[scale=0.5]{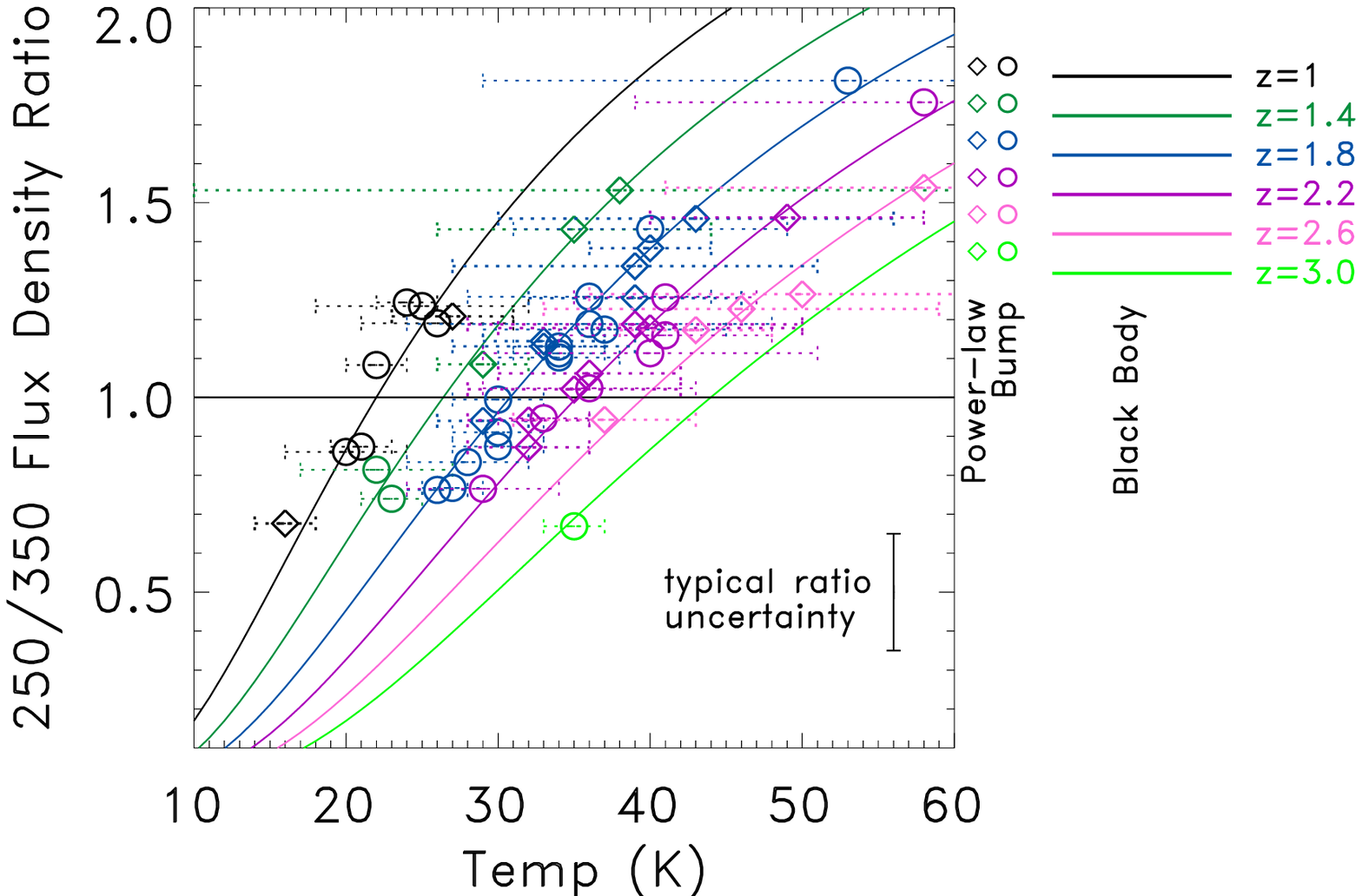}
\caption{\label{fig:tempratio} The observed-frame 250/350 \um\ ratio plotted as a function of redshift and temperature for single temperature modified black body models (lines, $S_\nu = B_\nu (T) *\nu^{1.5}$) and the DOGs (points).  We use the models and the observed 250/350 \um\ ratios of the DOGs to determine the characteristic far-IR dust temperatures of the galaxies.  DOGs with low 230/350 \um\ ratios tend to have cold dust temperatures, whereas galaxies with large 250/350 \um\ ratios tend to have warm temperatures.  These trends are modulated by redshift as the peak in the FIR dust emission shifts through the \Herschel\ passbands.  }
\end{figure}

The \Herschel\ SPIRE observations sample the far-IR SEDs of the DOGs near to the dust emission peak at rest wavelengths of $80-100$ \um.  Assuming the dust emission follows a simple black body, the 250/350 \um\ flux density ratio yields a characteristic temperature for the far-IR emitting dust peak \citep[e.g.][]{Dunne00, Draine03, Bussmann09b}.  To determine the far-IR dust temperature we construct synthetic dust models given by:
\begin{equation}
S_\nu = B_\nu (T) *\nu^\beta,
\end{equation}
where $B_\nu$ (T) is the black-body Planck curve and $\beta$ is the dust emissivity.  For this study, we assume a typical emissivity value of $\beta=1.5$ \citep[e.g.][]{Draine03}, and create 90 template spectra each with a different temperature ranging from $10-100$ K.  These synthetic spectra are then sampled at the SPIRE wavelengths, shifted to account for the redshifts of each DOG.  A fit between the model 250/350 \um\ flux density ratios with the actual data (Figure \ref{fig:tempratio}), reveals a characteristic far-IR temperature for each DOG.  Uncertainties on the temperatures, are estimated by altering the 250/350 \um\ ratios by their photometric uncertainties and recalculating the temperature.  The measured 250/350 \um\ ratios as a function of redshift and dust temperature are shown in Figure \ref{fig:tempratio},  for the 56 DOGs that were detected in both bands.

Figure \ref{fig:temp} shows histograms of the measured far-IR dust temperatures for the power-law and bump DOGs. The temperatures range from 19 K to 58 K.  However, the bulk of the temperatures are between 20 and 40 K.  In fact, the four DOGs with far-IR dust temperatures measured to be above 50K all have large temperature uncertainties, meaning that their temperatures are not significantly different from the larger sample.  Overall, there appears to be a trend of increasing dust temperature with increasing IR luminosity.  A similar trend is seen in local ULIRGs \citep[e.g.][]{Armus07}. However, as will be discussed in the following section, this trend may be partially the result of the \Herschel\ detection limits which vary with dust temperature.  

\begin{figure}
\includegraphics[scale=0.5]{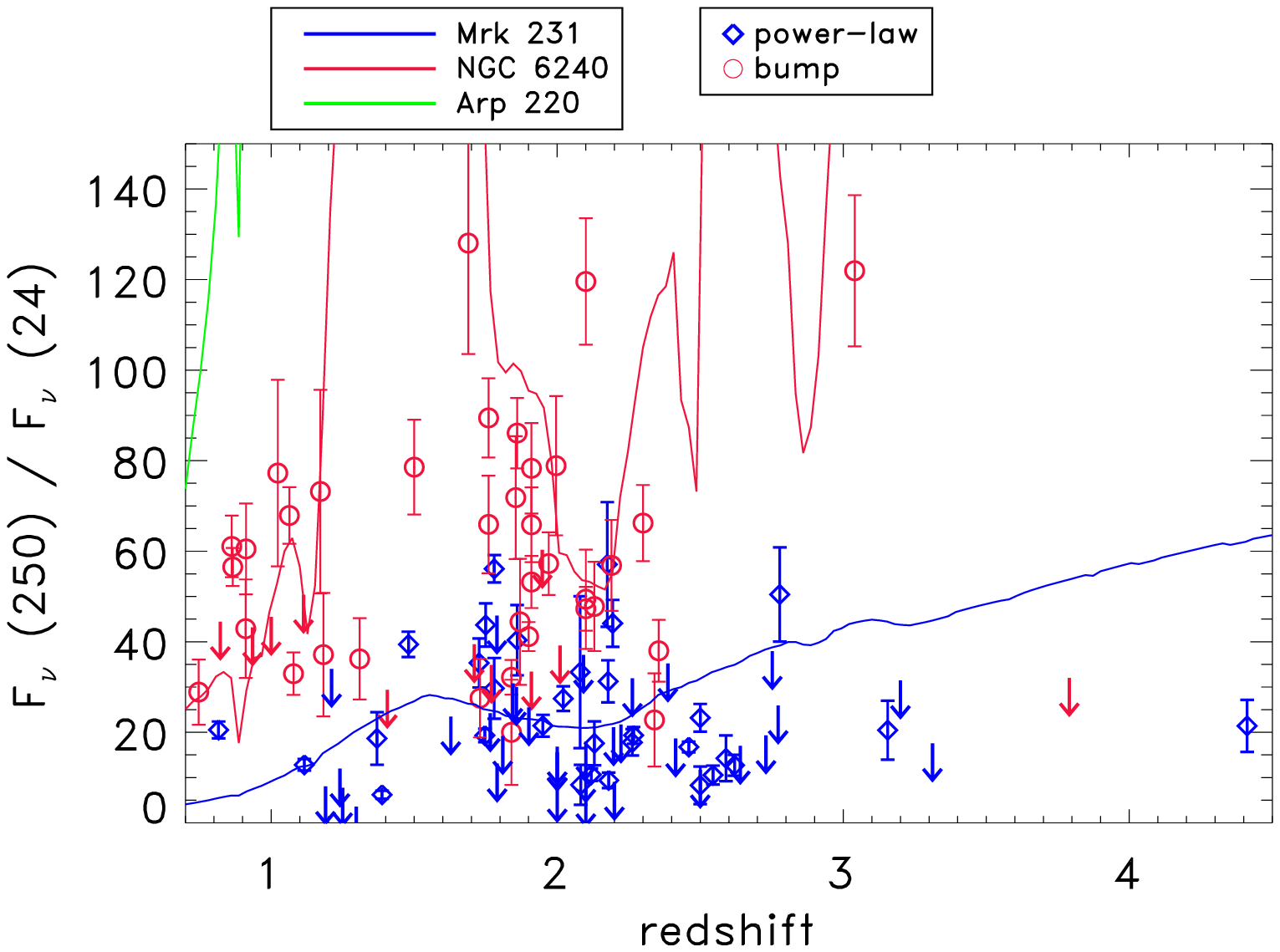}
\caption{\label{fig:250to24} The 250/24 \um\ flux density ratio plotted as a function of redshift for the DOG sample with spectroscopic redshifts. The mid-to-far-IR SED classifications of the DOGs are being driven by this ratio. The power-law DOGs (diamonds) have low 250/24 \um\ ratios compared with the bump DOGs (circles).  Over-plotted are the  250/24 \um\ flux density ratios for the local ULIRG templates, shifted with redshift. Power-law DOGs tend to follow the Mrk~231 ratios, while the bump DOGs tend to follow the NGC~6240 ratios (especially below redshift 3.5).  
Upper limits (arrows) for the DOGs not detected in \Herschel\ are also shown.  The limits are not radically different from the detected source ratios but are at the low end of the distributions suggesting that the typical ratio may be different for \Herschel\ detected vs. undetected sources.}
\end{figure}

The power-law and bump DOGs span a similar range of dust temperatures but the median dust temperature of the bump DOGs is lower than the median temperature of the power-law DOGs.  Both samples appear to be significantly cooler than a complete sample of local ULIRGs \citep[from the IRAS Bright Galaxy Sample; see ][]{Soifer87, Armus07} measured in the same way at the same rest-frame wavelengths. To estimate temperatures of the local ULIRGs, we redshift their SEDs to $z=2$, then observe them in the \Herschel\ SPIRE bands, determining their 250/350 \um\ flux density ratio in the same way as the high-$z$ galaxies.

The temperature measurements are given Table \ref{tab:photometry} and will be discussed further in the following section.  
            
\section{Discussion}
With the deep \Herschel\ SPIRE observations of the \boot\ field from HerMES, we can, for the first time, constrain the far-IR SEDs and hence the total \lir\ of large samples of $z=2$ DOGs.  This paper presents results for a sample 113 DOGs with spectroscopic redshifts, selected to have very high mid-IR-to-optical flux ratios.  In this sample, DOGs that show AGN like signatures in the rest-frame near-IR (power-law DOGs) tend to show AGN-like mid-to-far-IR SEDs. Meanwhile DOGs with starburst-like signatures in the rest-frame near-IR (bump DOGs) tend to show starburst-like SEDs at longer wavelengths. While the power-law DOGs are less likely to be detected at 250 \um, those that are detected are likely to have significantly higher IR luminosity.  

The discussion of these results, below, starts with a comparison of  the \Herschel\ far-IR photometry with other far-IR observations  of the sample galaxies.  Next, the detection biases of the SPIRE data, including both temperature and luminosity biases are discussed in detail. Then, we discuss the value of the mid-IR data from \Spitzer\ for accurate predictions of the IR luminosities of the DOGs.  Finally, the DOGs are compared with other high-$z$ samples of ULIRGs.  

\begin{figure}
\includegraphics[scale=0.5]{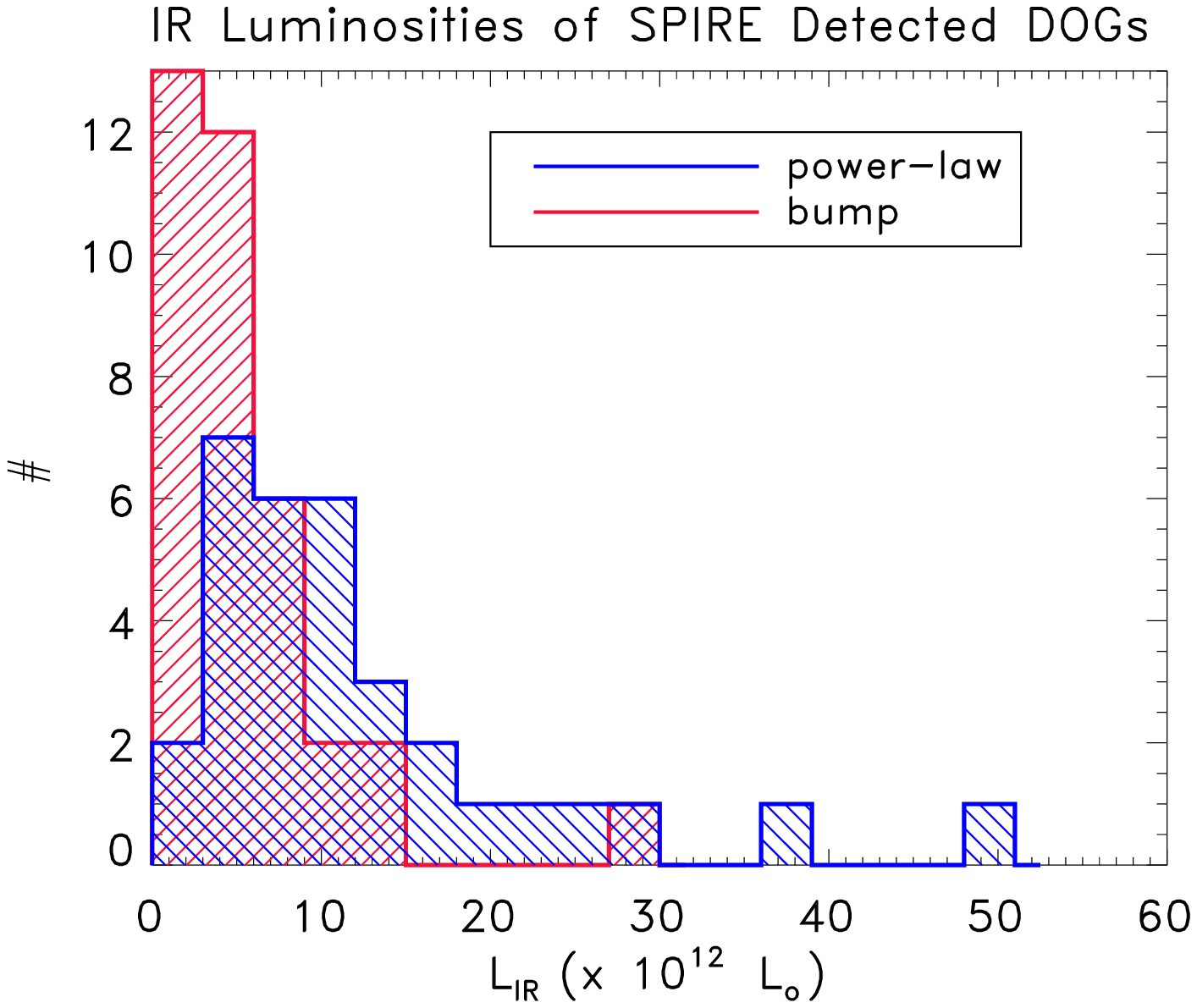}
\caption{\label{fig:LIR} \lir\ (8-1000 \um) measurements for the power-law (blue) and bump (red) DOGs as estimated from a simple interpolation of the SED in $F_\lambda$ vs. $\lambda$ space.  While the power-law DOGs are less likely to be detected at the SPIRE wavelengths, when they are detected, their \lir 's are typically higher than for the bump DOGs.  While the typical \Herschel-detected bump DOG is a ULIRG with \lir\ $< 10^{13}$ \lsun, $\sim50$\% of the \Herschel-detected power-law DOGs have higher IR luminosities, e.g. \lir\ $> 10^{13}$ \lsun. 
}
\end{figure}

\subsection{Comparisons with Previous Far-IR Observations of Our Sample}

Previously, 12 of the DOGs in the sample were observed at the Caltech Sub-mm Observatory with SHARC-II at 350 \um\ \citep{Bussmann09b}.  Only 4 were detected, while upper limits were derived for the remainder of the sample.  The \Herschel\ photometry are in good agreement with the previous results, returning fluxes below the SHARC-II detection limits, and roughly matching (within 1-2 sigma) the fluxes of the DOGs that SHARC-II did detect.  The  SHARC-II sample targeted several of the brightest 24 \um\ sources, which are predominantly power-law DOGs.  As we have shown, these sources generally have low 350/24 \um\ flux density ratios, and therefore are difficult to detect at 350 \um.  Sub-mm programs targeting 24 \um\ bright bump sources have generally shown a higher detection rate \citep[e.g.,][]{Lonsdale09, Kovacs10, Chapman10}, as expected, given their propensity for higher 350/24 \um\ flux density ratios.

Several of the DOGs in our sample were also previously detected at 70 and 160 \um\ with deep \Spitzer\ MIPS images  \citep{Tyler09}.  These observations constrain the blue side of the far-IR dust peak.  Seven DOGs were detected in 70 \um\ band while 10 were detected in the 160 \um\ band.  From these observations \citet{Tyler09} calculated \lir\ for 11 sources. The new SPIRE derived estimates of \lir\  agree with the Taylor estimates to within 20\%, which is quite good considering the potentially large systematic uncertainties. 

\subsection{Luminosity and Temperature Selection Biases of \Herschel\ Samples}
\begin{figure}
\centering
\includegraphics[scale=0.5]{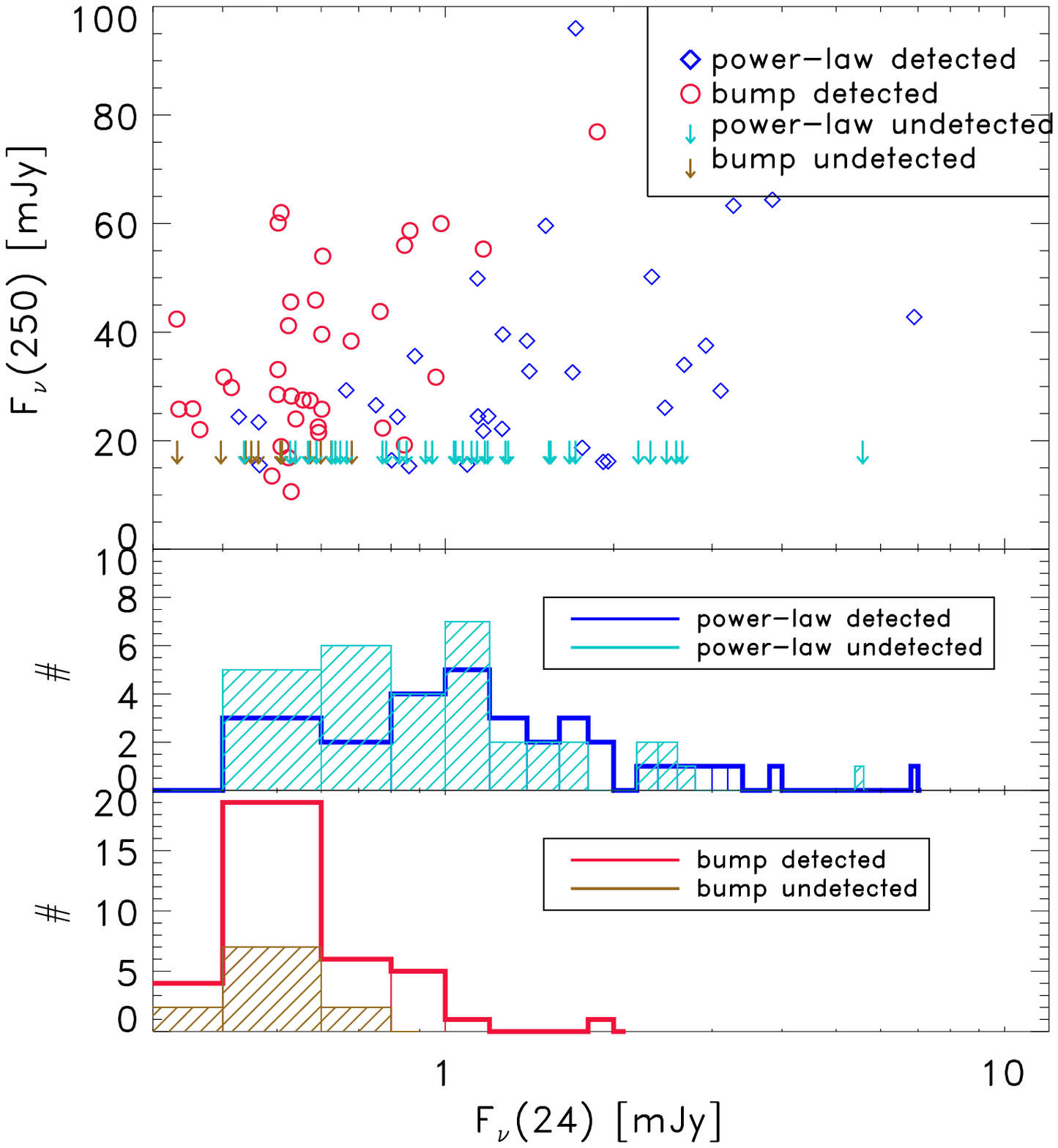}
\caption{\label{fig:250selected} Top: $F_\nu(250)$ plotted against $F_\nu(24)$ for the bump (circles) and power-law (diamonds) DOGs.  Also shown are 250 \um\ flux density upper limits for the objects not detected in \Herschel\ (arrows).  Middle: Histograms of the distribution of 24 \um\ flux densities for power-law DOGs that are detected (blue) and undetected (cyan) in \Herschel.  Bottom:  Histograms of the distribution of 24 \um\ flux densities for bump DOGs that are detected (red) and undetected (brown) in \Herschel.  The power-law DOGs show a much stronger dependence on 24 \um\ flux density for detection in \Herschel\ than the bump DOGs. Power-law DOGs fainter then $F_\nu (24) < 1$ mJy are only detected $\sim30$\% of the time, whereas, the bulk of the bump DOGs have $F_\nu (24) < 1$ and most are detected.}
\end{figure}

One of the surprising results from our study is that while the bump DOGs are more likely to have detections at SPIRE wavelengths (see Figure \ref{fig:FIRclass}), the power-law DOGs that are detected are likely to have higher \lir 's (see Figure \ref{fig:LIR}).  Selection biases   summarized in Figure \ref{fig:250selected} may be playing a role in these results. This figure compares the 24 \um\ flux densities of those DOGs that are detected at 250 \um\ with those that are undetected.  While the bulk of the bump DOGs with $F_\nu$ (24) $<1 $ [mJy] are detected at 250 \um, less than 50\% of the power-law DOGs are detected.  This is not surprising as the 250/24 \um\ ratio is small for the power-law DOGs and typically much larger for the bump DOGs. 

\begin{figure}
\centering
\includegraphics[scale=0.45,trim=0 0 0 50]{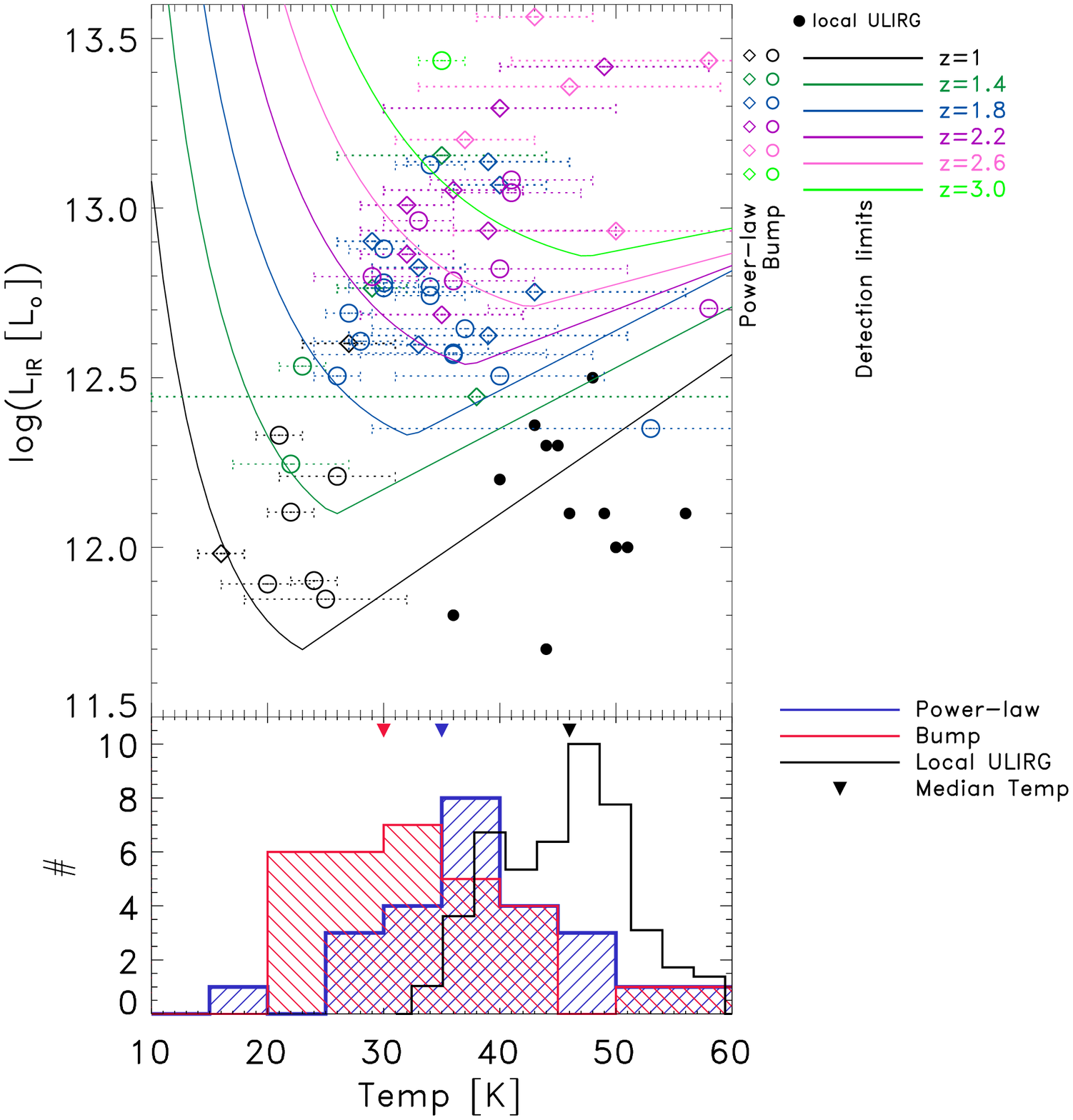}
\caption{\label{fig:temp} Top: IR luminosity plotted as a function of far-IR temperature for the power-law (diamonds) and bump (circle) DOGs and a complete sample of local ULIRGs from the IRAS bright galaxy survey (black points). Also included are the 3 sigma luminosity detection limits for the SPIRE 250 \um\ image as a function of temperature and redshift (colored lines). Bottom: the distribution of the measured far-IR dust temperatures of the power-law (blue) and bump (red) DOGs, compared with local ULIRGs (black) measured at roughly the same rest-frame wavelengths.  The DOGs that are detected in \Herschel\ tend to have cooler median far-IR dust temperatures (downward triangles) than the local ULIRGs, and the median temperature of the bump DOGs is about 5 K cooler than the power-law DOGs.  There is a general trend of increasing temperature with increasing IR luminosity. However this may be at least partially set by the detection limits of the sample which create biasses against detection at both the cold and warm ends of the distribution. For instance local ULIRGs would not be detected above $z=1.4$, because their dust temperature is too warm.}
\end{figure}

When a power-law DOG is detected at 250 \um\ it tends to have a larger 24 \um\ flux density for a given 250 \um\ flux density compared with the bump DOGs.  Therefore SPIRE-detected power-law DOGs will be more IR luminous (on average) than the bump sources. 

However, luminosity may not be the only selection bias in the \Herschel\ data.  Another bias to consider, is the temperature of the far-IR emitting dust \citep[e.g.][]{Chapman04, Chapman05, Pope06, Casey09, Symeonidis11}.  Figure \ref{fig:temp}, plots the IR luminosity of the DOGs as a function of the far-IR dust temperature.  For \Herschel\ detected DOGs, galaxies with higher IR luminosities tend to have warmer dust temperatures.  This result can be explained at least in part by the SPIRE detection limits for galaxies of a given temperature and IR luminosity (colored lines).  The warm-side limits were generated by scaling the SEDs of a complete set of 12 local ULIRG (which span a range of temperatures from $35 -60$ K) to different IR luminosities, and  ``observing'' them at high-$z$ in the \Herschel\ bands.  We then determined the luminosity at which they would be detected in the SPIRE 250 \um\ band (to a 20 mJy limit), at the same rest wavelength as the DOGs as a function of redshift.  For instance, at $z=1$  any 20 K ULIRGs will be detected in SPIRE observations of \boot, but only the most IR luminous (e.g., \lir $>10^{12.6}$ \lsun) 50 K ULIRGs will be detected.  None of the local ULIRGs would actually be detected in SPIRE if they were above $z=1.4$. The cold temperature detection limits ($T<30$ K) were generated in a similar way with  modified black body spectra.  As can be seen in Figure \ref{fig:temp}, there are also strong selection biases against detecting very cold sources with SPIRE.        

The temperature bias requires that objects with warmer dust must have higher IR luminosities to be detected in SPIRE. Thus, the \Herschel\ non-detected sources could be missed because they are lower luminosity, have a warmer temperature, or both. However, above \lir $=10^{13}$ \lsun, even the warm objects (40--60 K) should be detected regardless of redshift (Figure \ref{fig:temp}).  Therefore, the undetected DOGs, including the 51\% of the power-law sources that are not detected, must have \lir $<10^{13}$ \lsun. 

\subsection{Predicting \lir\ from 24 \um\ Flux Density}


Figure \ref{fig:LIRcomp1} shows  the \lir / $\nu L_\nu$(24 \um\ observed-frame) ratio as a function of redshift.  For the power-law DOGs, the \lir / $\nu L_\nu$(24) values lie in a fairly tight range of $6.5\pm1.4$. This suggests that the 24 \um\ luminosity can be used to predict the IR luminosities of the power-law DOGs to within roughly 20\%.  

The scatter in Figure \ref{fig:LIRcomp1} is significantly larger for the bump DOGs so a similarly simple prediction is not possible for their \lir 's.  However, the flux-dependent relation predicted by \citet{Chary01} appears to predict \lir\ for the bump DOGs with reasonable accuracy. Figure \ref{fig:LIRcomp2} compares the \Herschel\ derived \lir 's of the DOGs to the predicted values from the templates of \citet{Chary01}.  For the bump DOGs, these relations work across the full range of \lir 's.  Not surprisingly, these relations tend to over-predict the \lir 's of the power-law DOGs, as they were designed for star forming galaxies, not obscured AGN, which have larger 24 \um\ contributions from warm dust. However, even for the power-law DOGs the \citet{Chary01} relations are good to within 50\%. 

The fact that the \citet{Chary01} templates work so well for the bump DOGs is somewhat surprising because these relations have been shown to fail for other samples of optically-bright high-z ULIRGs \citep[e.g.][]{Pope06, Muzzin10, Elbaz11, Rujopakarn11}.  These other studies find that most high-z ULIRGs are just scaled up versions of local star forming galaxies rather than having far-IR SEDs similar to local ULIRGs.  The situation is reversed for the bump DOGs,  local ULIRG templates are a good match to the 250/24 \um\ flux density ratios and hence IR luminosities of the bump DOGs.  

While these relations work for the sources that are detected in the SPIRE data, they may not work for the DOGs that are not detected in SPIRE, especially the large numbers of undetected power-law DOGs.  In order for these relations to work more generally, the undetected DOGs must have similar 250/24 \um\ ratios as the detected DOGs. The upper limits on the 250/24 \um\ flux density ratios of SPIRE-undetected DOGs (shown in Figure \ref{fig:250to24}) tend to be at or below the flux ratios for the detected DOGs at a given redshift. However, the limits are not dramatically lower than the flux densities for the detected sources. 

A simple stacking analysis on the  \Herschel\ images of the undetected DOGs gives a mean 250/24 \um\ flux density ratio of $7.8\pm1.3$ for the power-law DOGs, and $21.3\pm4.6$ for the bump DOGs.  (To get sufficient statistics we needed to bin across the full redshift range for the two sample types.)  As with the limits, these values are at the low end of the distributions of 250/24 \um\ flux density ratios of the SPIRE detected DOGs.  Thus we may be seeing evidence for a modest change in the mid-to-FIR SED shape for some DOGs.  It is not clear if this change is purely a luminosity effect, with the undetected sample having lower total \lir\ for a given 24 \um\ flux density, or if this change is a far-IR temperature effect, with the undetected DOGs possibly lacking a large reservoir of the coldest dust. That being said, the simple relations for estimating \lir\ given above are likely to be off by only modest amounts, as the detected sources with low 250/24 \um\ flux density ratio have measured \lir 's to within 50\% of their \citet{Chary01} predicted values.  This agreement is far superior to the previous uncertainties on \lir\ for the DOGs which exceeded factors of two \citep{Dey08}.   


 
\begin{figure*}[t]
\includegraphics[scale=0.55]{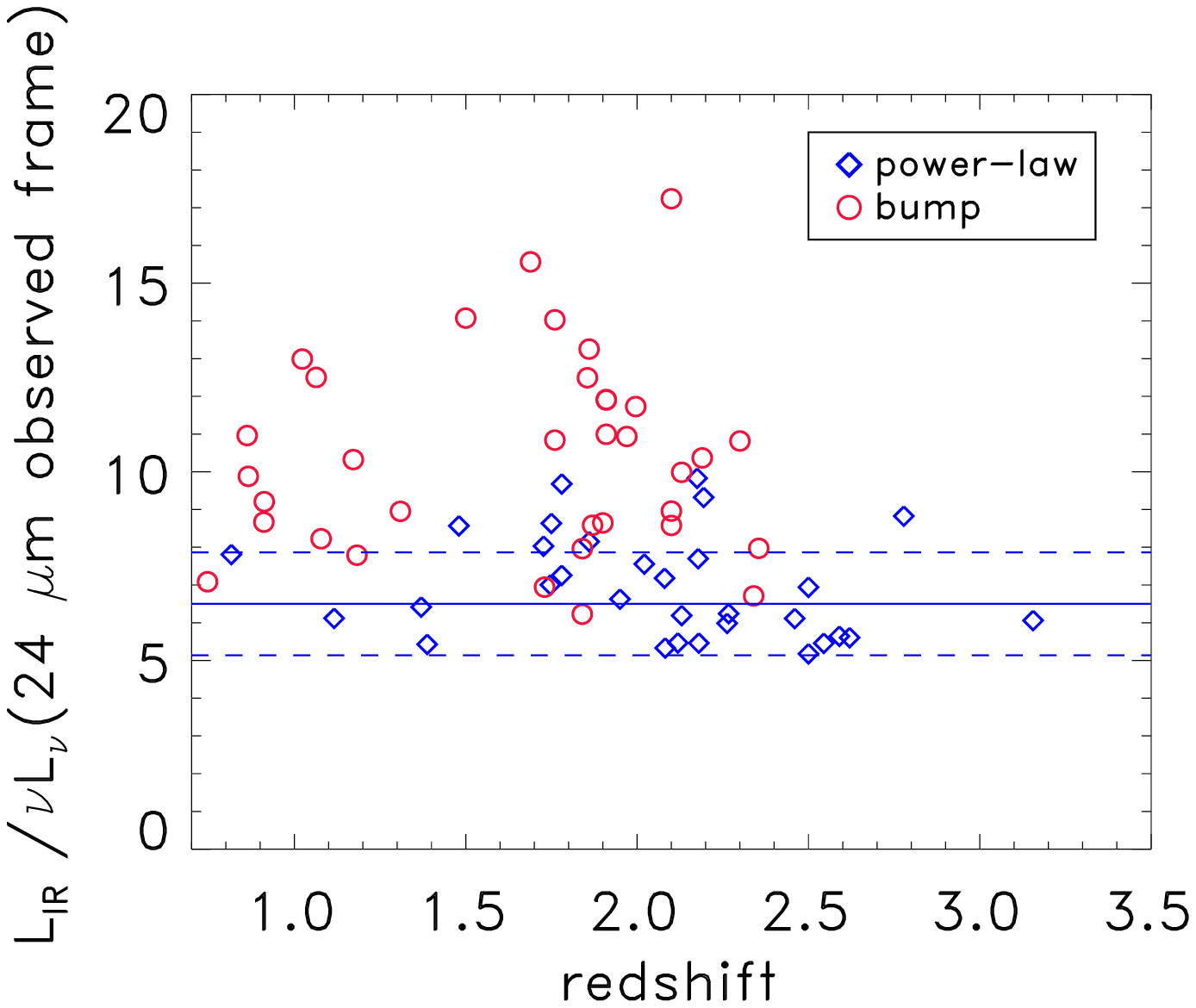}
\includegraphics[scale=0.55]{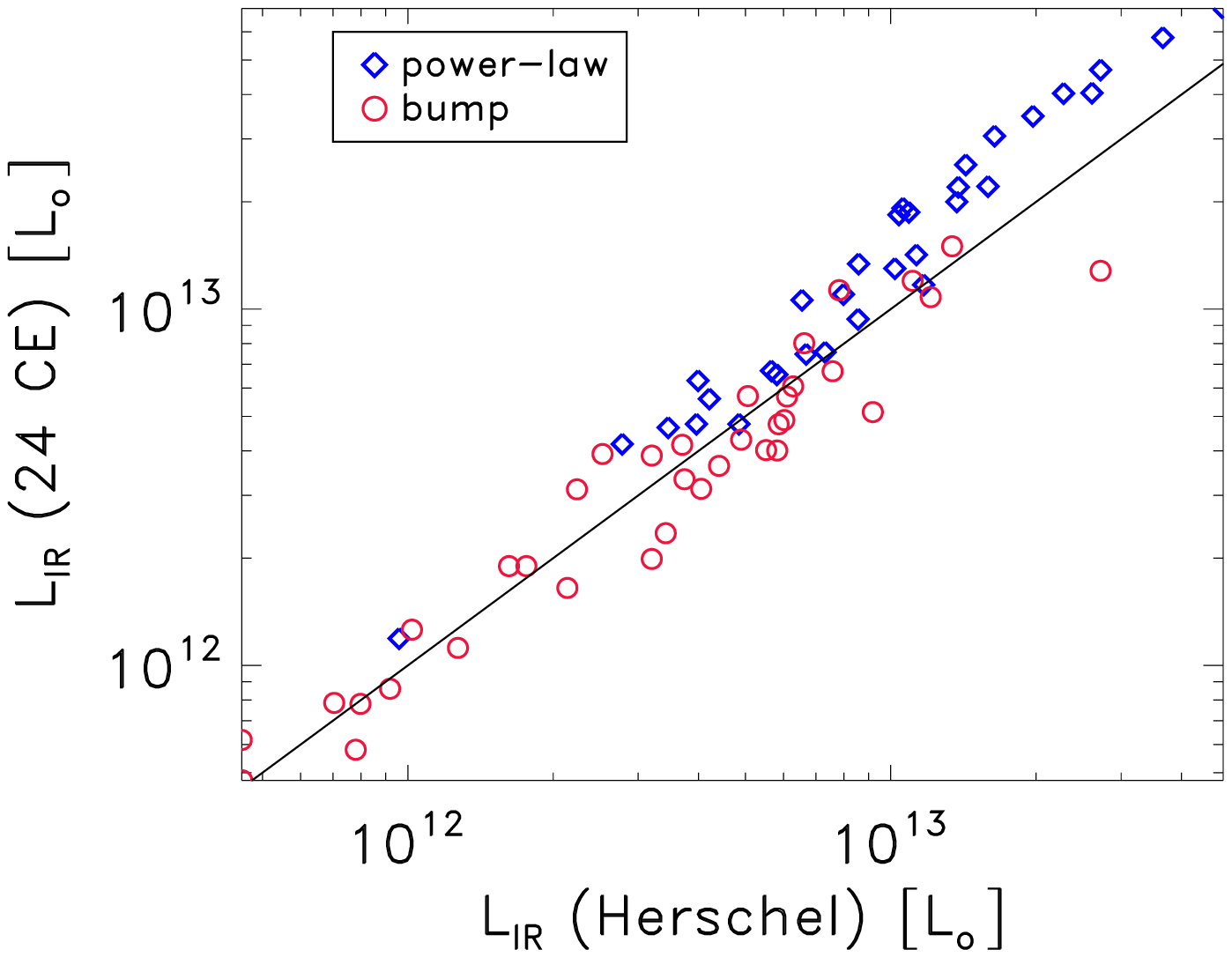}
\caption{\label{fig:LIRcomp} Left: $L_{IR} / \nu L_\nu(24)$ (observed frame) plotted as a function of redshift for the power-law (blue diamonds) and bump (red circles) DOGs.   For the power-law DOGs, \lir\ is well predicted by 24 \um\ luminosity  with a mean $L_{IR} / \nu L_\nu (24) = 6.5\pm1.4$.  The much larger scatter of the bump DOGs, especially around $z=2$ when the 8 \um\ PAH features shift into the 24 \um\ passband, means that a simple relation will not work well for predicting \lir's of the bumps. Right: The \citet{Chary01} predicted IR luminosity, based on the 24 \um\ flux density and redshift, plotted as a function of the measured IR luminosity from the \Spitzer\ and \Herschel\ photometry. The flux dependent relation from \citet{Chary01} works quite well for predicting the true IR luminosity of the bump DOGs.  However, \citet{Chary01} has been shown to fail for other samples of $z=2$ ULIRGs which behave more like scaled up versions of local star forming galaxies \citep{Elbaz11}.  The \citet{Chary01} templates tend to over-predict the \lir 's of the power-law DOGs because these galaxies have an excess of warm dust from the central AGN.}
\end{figure*}

\subsection{Comparing the DOGs to Other Galaxy Samples}

\begin{figure}
\centering
\includegraphics[scale=0.5,trim=0 0 0 40]{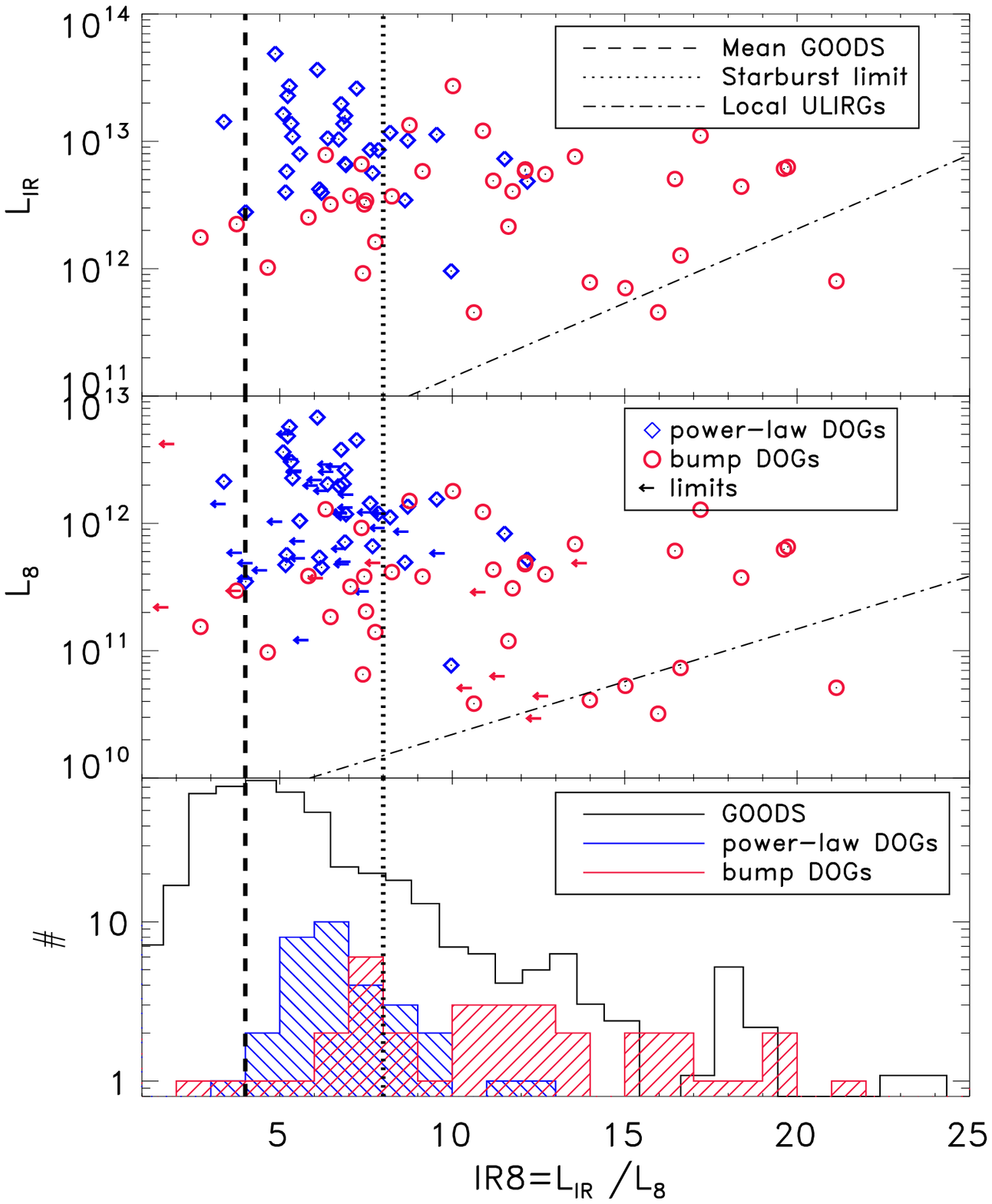}
\caption{\label{fig:IR8} Top: IR luminosity plotted as a function of $IR8=L_{IR}/L_8$, where $L_8=\nu L_\nu (8$\um\ rest-frame).  Power-law DOGs are shown as blue diamonds, while bump DOGs are shown as red circles.  The median $IR8$ value, for $z=1-2$ LIRGs and ULIRGs in GOODS, is shown as the thick vertical dashed line.  The division between ``main sequence'' and ``starburst'' galaxies is shown by the thick vertical dotted line, with starburst galaxies exhibiting higher $IR8$ values. The \lir\ vs. $IR8$ for local ULIRGs is shown as the dot-dashed line, which lies in the starburst region.  Middle: Same as top only now $L_8$ is plotted as a function of $IR8$.  Limits on $IR8$ for \Herschel\ non-detected galaxies are shown as arrows.  Bottom: Histograms of $IR8$  values for galaxies in the GOODS field \citep[black,][]{Elbaz11}, compared with the power-law (blue) and bump (red) DOGs. The bulk of the GOODS galaxies including the typical high-z LIRGs and ULIRGs have $IR8 \sim4$ (dashed line), defining a main sequence of star formation at $z=1-2$.  Galaxies with high $IR8$ values ($> 8$, dotted line) are assumed to be in a starburst mode with star formation occurring in very high density gas where PAH emission is suppressed compared to \lir.  The power-law DOGs have tight distribution of $IR8$ values with a mean around $IR8\sim6$.  Meanwhile the bump DOGs show a wide range of $IR8$ values, however, most are high compared with the average $z=2$ LIRGs and ULIRGs in GOODS (e.g., dashed line).  Bump DOGs have $IR8$ values similar to local ULIRGs and high-$z$ starburst rather than like main-sequence $z=2$ ULIRGs. }
\end{figure}

\citet{Elbaz11} presents the \Herschel\ derived far-IR SEDs of star forming galaxies in the GOODS fields.  They find that the bulk of them, including the $z=1-2$ LIRGs and ULIRGs, follow an infrared main sequence which they define based on the ``$IR8$'' parameter, where $IR8= L_{IR} /L_{8}$ and $L_{8} = \nu L_\nu (8$\um) is the luminosity at rest-frame 8 \um. $L_8$, is a good proxy for the PAH emission strength from star formation.  For most star forming galaxies in the local universe, PAH strength tracks \lir\ in a predictable fashion, e.g. $IR8 = L_{IR} / L_8 \sim4$ \citep{Elbaz11}.  These normal star forming galaxies define the infrared main sequence and also show a tight range of specific star formation rates \citep[see for instance, ][]{Noeske07, Elbaz07, Daddi07, Daddi09, Pannella09, Magdis10}.  However, for galaxies undergoing a rapid starburst, PAH strength no longer tracks \lir, and $IR8$ increases.  In the local universe, ULIRGs typically lie off of the IR main sequence.  They have $IR8 >> 4$ (Figure \ref{fig:IR8} shows the local sample assembled in Elbaz et al. 2011, drawn from {\emph{AKARI}}, {\emph{ISO}}, and \Spitzer\ missions).  At $z=1-2$, however, \citet{Elbaz11} find that most LIRGs and ULIRGs not only have scaled up \lir\ values but also scaled up PAH strength.  This suggests that the mode of star formation in the typical $z=1-2$ LIRGs and ULIRGs is more similar to local star forming galaxies than it is to local ULIRGs, and that selecting on \lir\ alone is not a good way to isolate extreme star-bursting galaxies.

To compare the DOGs with these other samples, we calculate $IR8$ values for all of our sample galaxies detected in \Herschel.  For the bump DOGs we use a scaled version of the NGC~6240 template to estimate $L_8$, and for the power-law DOGs we use a scaled  up Mrk~231 template.  We scale the template to match the observed 24 \um\ flux of the DOG.  Then, as was done by Elbaz et al.,  we measure the mean flux density at rest-frame 8 \um\ within a ``filter'' that matches the \Spitzer\ IRAC 8 \um\ filter (i.e. channel 4).   We then convert to $L_8$ using the luminosity distance.  
 
Figure \ref{fig:IR8} compares the $IR8$ values from \citet{Elbaz11} with those of the DOGs. As described above, the bulk of the GOODS galaxies lie in a tight range of $1 < IR8 < 8$, with a peak at $IR8=4$.  The GOODS-sample does contain a tail of galaxies with $IR8 > 8$ which are classified as burst mode galaxies.  
In contrast with the typical GOODS galaxies, the median $IR8$ values of the DOGs are significantly higher.  The power-law DOGs show a tight distribution centered on $IR8\sim6$. We saw this same result in the previous section where we found  \lir / $\nu L_\nu$(24 \um\ observed frame) $=6.5\pm1.4$.  In contrast, the bump DOGs show an wide range in $IR8$, but prefer high values.  Only a handful are near the peak of the normal GOODS galaxies of $IR8=4$.  The $IR8$ values of the bump DOGs are closer to those of the local ULIRGs (which also have high $IR8$ values) and the star-bursting samples in GOODS, rather than the main sequence $z=2$ ULIRGs. They also overlap with sub-mm galaxies which typically have even higher $IR8\sim20$ \citep{Pope08a}.    

For those 11 bump DOGs observed with \Spitzer\ IRS \citep{Desai09} we directly measured $L_8$ from the spectrum.  All of the IRS derived $IR8$ values match those derived from NGC~6240 to within better than 50\%, and in no cases do the IRS derived values change whether a DOG would be in the starburst vs. main sequence region of the $IR8$ plot. For 7 of the 11 bump DOGs observed with IRS, the IRS derived $IR8$ values are higher than those derived from NGC~6240.  

\begin{figure}
\centering
\includegraphics[scale=0.5]{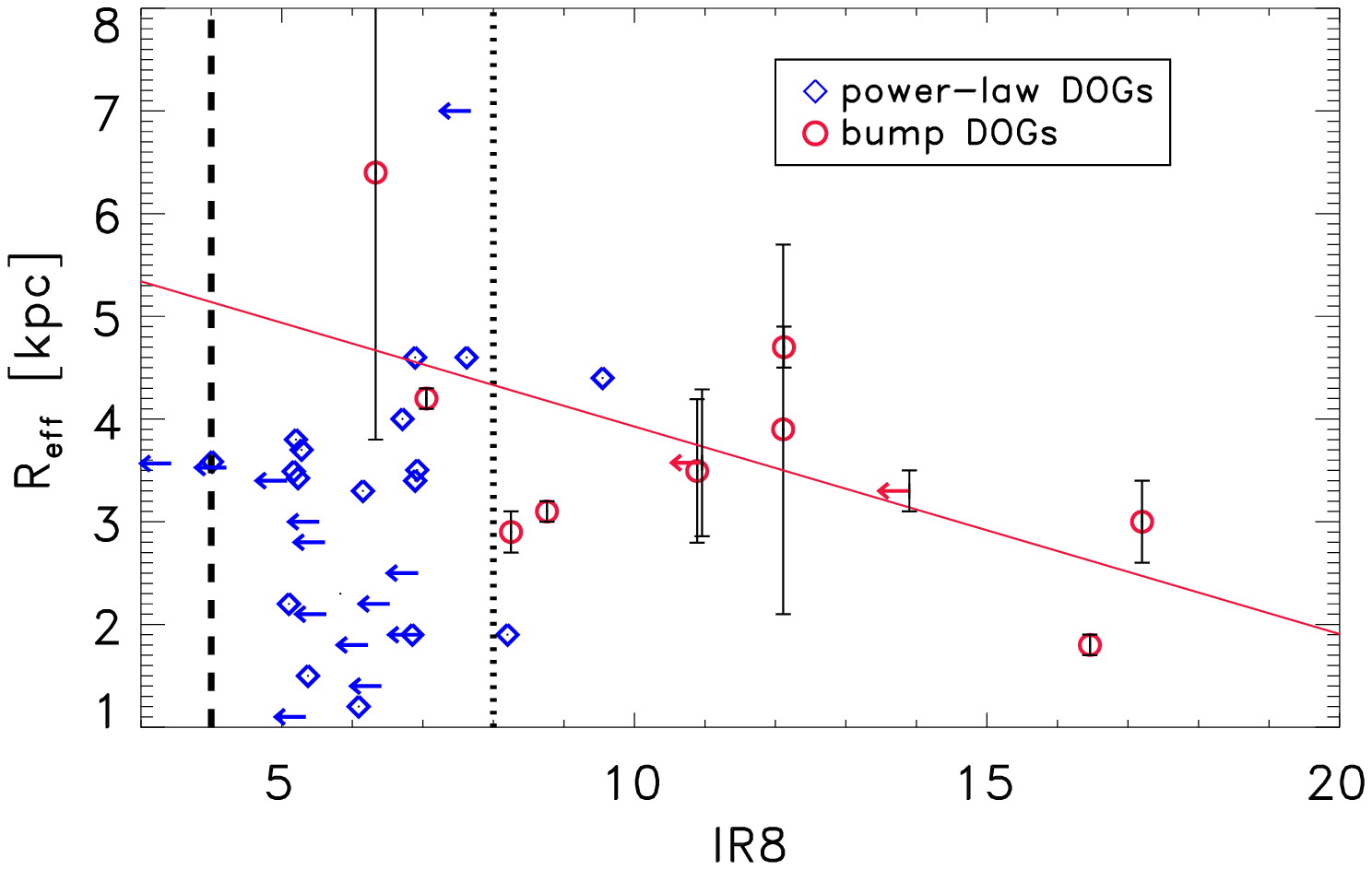}
\caption{\label{fig:IR8size} Morphological half-light radius plotted as a function of $IR8$ for the power-law (diamonds) and bump (circles) DOGs.  $IR8$ limits for \Herschel\ non-detected DOGs are also shown (arrows).  The median $IR8$ value, for $z=1-2$ LIRGs and ULIRGs in GOODS, is shown as the thick vertical dashed line.  The division between ``main sequence'' and ``starburst'' galaxies is shown by the thick vertical dotted line, with starburst galaxies exhibiting higher $IR8$ values.  Size is measured from rest-frame optical light in \HST\ NICMOS \citep{Bussmann09, Bussmann11} or Keck AO imaging \citep{Melbourne09}.  While the power-law DOGs show no obvious trend of $IR8$ with size, there is a correlation between the two for bump sources (red line, with a Pearson correlation coefficient, $\rho = -0.62$).  The most compact  objects tend to have the highest $IR8$ values.  This is similar to what \citet{Elbaz11} found for local ULIRGs only they were able to measure size in the mid-IR.  Compact sizes may decrease PAH to total IR emission in both the local and high-$z$ ULIRGs. For the local sample the most compact sources have undergone a recent merger.  }
\end{figure}

\citet{Elbaz11} points out that $IR8$ values tend to increase when the star formation is occurring in morphologically  compact regions.   In the local ULIRGs, these highly compact star forming regions are typically the result of major mergers funneling gas to the centers of these systems. It is not clear if the same merger related processes are leading to the high $IR8$ values of the DOGs.  While there is certainly evidence for some merging in the DOG samples \citep{Melbourne09, Bussmann09, Donley10}, the fractions with obvious major merger signatures remain small, less than 30\%. 

For the bump DOGs there does appear to be a trend of decreasing effective radius with increasing $IR8$ value, as shown in Figure \ref{fig:IR8size}.  This result may be indicating that the high $IR8$ values of the DOGs are also associated with more compact geometry. We caution, however,  the sample with radius measurements is small. In addition, these sizes are measured from near-IR \HST\ \citep{Bussmann09, Bussmann11} and Keck AO \citep{Melbourne08b, Melbourne09} images of the DOGs, and therefore trace the stellar light rather than the star forming gas.  A better comparison would be to determine the characteristic sizes of the star forming gas itself, for instance with the Atacama Large Millimeter Array (ALMA).    


The moderately high $IR8$ values of the power-law DOGs also differentiate them from the lower luminosity AGN in GOODS.  \citet{Elbaz11} shows that both the x-ray selected and IR selected AGN in GOODS tend to follow the same IR8 trend lines (i.e. $IR8\sim4$) as the non-AGN systems.  In contrast the power-law DOGs prefer somewhat higher $IR8$ values ($IR8\sim6$).  This basically means that for a given amount of rest-frame 8 \um\ flux the power-law DOGs have higher IR luminosities than the typical GOODS AGN.   The power-law DOGs could have higher fractions of cold dust than the GOODS AGN, which would tend to increase \lir\ without increasing $L_8$, or they could just be producing more IR luminosity for a given amount of PAH emission. 


Some star formation, even in the power-law DOGs, would not be a major surprise.  For instance, \citet{Mullaney11} found that the X-ray selected AGN in GOODS have far-IR SEDs very similar to normal star forming galaxies and are likely to have ongoing star formation.  Likewise, while the IR luminosity of Mrk~231 is dominated by hot dust from an AGN, there is strong evidence for significant circum-nuclear star formation of as much as 100 \msun \yr\ \citep{Davies04}. Thus the power-law DOGs, which have SEDs similar to Mrk~231, may also host some star formation.  This additional star formation could increase $IR8$ if it is also in a low PAH mode.  

Again, the DOGs that are not detected in \Herschel\ may behave differently in the $IR8$ plots from the detected ones.  However, their $IR8$ limits do not suggest significantly lower $IR8$ values (see Figures \ref{fig:IR8} and \ref{fig:IR8size}), except for a handful of sources.  \citet{Pope08} showed that for 12 lower luminosity (\lir $\sim 1\times10^{12}$) DOGs in the GOODS field  that $IR8\simeq7$, so there may be some luminosity dependence on these results.

While the $IR8$ values and the observed-frame 250/24 \um\ ratios of the DOGs are similar to the local ULIRGs, their far-IR dust temperatures (as measured by the observed-frame 250/350 \um\ ratio) tend to be cooler.  The median temperature of the \Herschel -detected bump DOGs is 30 K, which is $10-20$ K degrees cooler than the local ULIRGs measured in the same way (see Section 3.3).  The dust temperatures of the \Herschel -detected power-law sources are only slightly higher (median $T=35$ K). In fact, the median far-IR temperature of the bump DOGs is also about 10 degrees cooler than the median temperature of the GOODS star-burst samples.  

Sub-mm galaxies also exhibit extreme star-formation rates and cold dust temps \citep[e.g.,][]{Chapman05, Kovacs06, Chapman10}.  The bump DOGs show very similar far-IR temperatures to the sub-mm galaxies. Thus, while the \Herschel-detected DOGs appear to primarily be scaled up versions of local ULIRGs, they also likely host additional cold dust not seen in local ULIRGs or the other high-$z$ starbursts, except for sub-mm galaxies. These results may suggest a deeper connection between bump DOGs and sub-mm galaxies, than was possible to make based on shorter wavelength data alone.

    


\section{Conclusions}
We use \Herschel\ SPIRE observations in the \boot\ field of the NDWFS, to constrain the far-IR SEDs of a sample of 113 optically faint $z=2$ ULIRGs selected to have R - [24] $>14$ [mag] (i.e., $F_{\nu}$(\tf)$ / F_{\nu} (R) \ga 1000$).  Galaxies selected this way are termed dust obscured galaxies or DOGs and are among the most luminous objects at $z=2$. 

We find that the observed-frame 250/24 \um\ flux density ratios of the \Herschel\ detected DOGs (60\% of the sample) are well predicted by their rest-frame  near-IR SEDS.   DOGs with power-law SEDs at near-IR wavelengths tend to have 250/24 \um\ ratios similar to the local AGN dominated ULIRG, Mrk~231.  DOGs with a stellar bump in their rest-frame near-IR SED tend to have 250/24 \um\ ratios similar to the local star-burst ULIRG, NGC~6240.  

The \lir's of the \Herschel\ detected DOGs are also well predicted from their fluxes at shorter wavelengths.  The IR luminosities of the bump DOGs are well predicted from the \citet{Chary01} templates that scale with 24 \um\ flux density. Power-law DOGs have \lir 's that are well predicted from an even simpler relation between their observed-frame 24 \um\ luminosity and IR luminosity,  \lir / $\nu L_\nu$(24) $=6.5\pm1.4$.  

Power-law DOG exhibit lower 250/24 \um\ flux density ratios than bump DOGs.  Therefore, those power-law DOGs that are detected in SPIRE typically have much higher 24 \um\ fluxes and \lir 's compared with bump DOGs at the same 250 \um\ flux. Indeed,  $\sim50$\% of the SPIRE detected power-law DOGs have \lir $>10^{13}$ \lsun, whereas the SPIRE detected bump DOGs typically have \lir $< 10^{13}$ \lsun.  The \Herschel\ detected power-law DOGs are likely to contain some cold dust (boosting the observed 250 \um\ flux densities) but their high IR luminosities are likely driven by the warm dust traced by the observed-frame 24 \um\ flux. In contrast the bump DOG luminosity is likely to be dominated by emission from cold dust. 


\citet{Elbaz11} finds that a large fraction of the $z=1-2$ LIRGs and ULIRGs in GOODS have $IR8=L_{IR}/\nu L_\nu(8$\um\ rest-frame$)\approx4$ placing them on the main-sequence of star forming galaxies at those redshifts. In contrast, the bump DOGs tend to have high $IR8$ values, i.e. $IR8 >>  4$, placing them in a star burst regime.  High $IR8$ values are more typical of starburst driven ULIRGs in the local universe, and of sub-mm galaxies at $z\sim2$, where star formation is occurring in very dense regions rather than in more spatially extended disks \citep{Elbaz11, Rodighiero11}.  We do find a trend whereby bump DOGs with smaller physical sizes (in stellar light) show higher $IR8$ values.  Additionally, while  other $z=2$ main sequence LIRGs and ULIRGs have 250/24 \um\ flux density ratios similar to lower luminosity local star forming galaxies \citep{Muzzin10,Elbaz11}, the DOGs have 250/24 \um\ flux density ratios well matched to local ULIRGs. 

However, the Herschel detected DOGs have cooler far-IR temperatures than local ULIRGs, $\sim30-40$ K as compared to the $40-50$ K for local ULIRGs.  The dust temperatures for the DOGs is quite similar to those found for sub-mm galaxies. Selection biases may play a role in the distribution of measured temperatures of the DOGs.  DOGs with warm far-IR dust temperatures need to have  significantly higher IR luminosities to be detected at SPIRE wavelengths compared with DOGs with cool far-IR dust temperatures. However, the large fraction that do have cool temperatures suggest that some DOGs harbor a cool gas reservoir, that can boost their far-IR flux.
  
There is some evidence (from detection limits and stacking) that the SEDs of the SPIRE-undetected DOGs exhibit lower observed-frame 250/24 \um\ ratios then the SPIRE-detected DOGs.  If these trends hold then the simple predictions of \lir\ given above may be over-estimated by a small factor ($<50$\%) for the far-IR faint DOGs.  Similarly, a lower 250/24 \um\ ratio would likely mean that the undetected DOGs have lower $IR8$ values than the SPIRE detected galaxies. Again, it is not clear if the non-detections are  the result of lower IR luminosity, higher far-IR dust temperature, or both.

\acknowledgments
This work is based (in part) on observations made with the \Spitzer\ Space Telescope, which is operated by the Jet Propulsion Laboratory, California Institute of Technology under a contract with NASA.  We would like to acknowledge the MIPS GTO team for producing the \Spitzer\ 24 \um\ imaging and source catalogues of the \boot\ field.  SPIRE has been developed by a consortium of institutes led by Cardiff University (UK) and including Univ. Lethbridge (Canada); NAOC (China); CEA, LAM (France); IFSI, Univ. Padua (Italy); IAC (Spain); Stockholm Observatory (Sweden); Imperial College London, RAL, UCL-MSSL, UKATC, Univ. Sussex (UK); and Caltech, JPL, NHSC, Univ. Colorado (USA). This development has been supported by national funding agencies: CSA (Canada); NAOC (China); CEA, CNES, CNRS (France); ASI (Italy); MCINN (Spain); SNSB (Sweden); STFC (UK); and NASA (USA).  We also would like to acknowledge the HerMES collaboration for providing this excellent dataset across the \boot\ field.    The US \Herschel\ Science Center also provided a workshop on SPIRE image reduction and photometry that was very valuable for our understanding of the data. The research activities of AD and BTJ are supported by NOAO, which is operated by the Association of Universities for Research in Astronomy (AURA) under a cooperative agreement with the National Science Foundation. Facilities used: \Herschel\ Space Telescope; \Spitzer\ Space Telescope; Kitt Peak National Observatory Mayall 4m telescope; W. M. Keck Observatory; Gemini-North Observatory
 
\bibliographystyle{/Users/jmel/bib/apj}
\bibliography{/Users/jmel/bib/bigbib2}

\clearpage
\LongTables

\begin{deluxetable}{lcccccc}
\tabletypesize{\scriptsize}
\tablecaption{Properties of the Sample \label{tab:sample}}
\tablehead{\colhead{Galaxy ID} & \colhead{RA} & \colhead{Dec} & \colhead{redshift} & \colhead{R-[24]} & \colhead{$F_\nu (24 \mu m) / F_\nu(R)$} & \colhead{$F_\nu (24 \mu m)$ [mJy]}}
\startdata
                        SST\_J142538.2+351855&           14 25 38.210&          +35 18 55.41 &  2.30& 17.60&2.56e+04&$  0.85\pm  0.05$\\
                        SST\_J142541.3+342420&           14 25 41.301&          +34 24 20.51 &  2.19& 14.45&1.40e+03&$  0.67\pm  0.04$\\
       SST\_J142544.5+344558\tablenotemark{a}&           14 25 44.586&          +34 45 58.32 &  0.75& 15.65&4.26e+03&$  0.77\pm  0.04$\\
                        SST\_J142554.9+341820&           14 25 54.994&          +34 18 20.80 &  4.41& 15.36&3.24e+03&$  1.14\pm  0.05$\\
                        SST\_J142607.8+330425&           14 26 07.855&          +33 04 25.79 &  2.09& 14.29&1.21e+03&$  0.54\pm  0.05$\\
                        SST\_J142622.0+345249&           14 26 22.031&          +34 52 49.47 &  2.00& 15.08&2.51e+03&$  1.30\pm  0.05$\\
                        SST\_J142626.4+344731&           14 26 26.499&          +34 47 31.20 &  2.10& 19.02&9.50e+04&$  1.17\pm  0.04$\\
                        SST\_J142637.3+333025&           14 26 37.397&          +33 30 25.65 &  3.20& 16.53&9.59e+03&$  0.64\pm  0.05$\\
\enddata
\tablenotetext{a}{DOGs not in Bussmann et al. 2011}
\end{deluxetable}

\clearpage
\begin{deluxetable}{lccccccc}
\tabletypesize{\scriptsize}
\setlength{\tabcolsep}{0.02in} 
\tablecaption{Multi-Wavelength Photometry (with limits in parentheses) \label{tab:bluephot}}
\tablehead{\colhead{Galaxy ID} & \colhead{$Bw$} & \colhead{$R$} & \colhead{$I$}  &\colhead{3.6} &\colhead{4.5} &\colhead{5.8} & \colhead{8.0} \\
 &  \colhead{[Vega mag]} & \colhead{[Vega mag]} &\colhead{[Vega mag]}  & \colhead{[$\mu$Jy]} &\colhead{[$\mu$Jy]} &\colhead{[$\mu$Jy]} &\colhead{[$\mu$Jy]}}
\startdata
SST\_J142538.2+351855&(27.1)&(26.1)&(25.4)&   19.35 $\pm$     2.44&   26.70 $\pm$     3.40&   30.89 $\pm$    10.21&   44.02 $\pm$     8.04\\
SST\_J142541.3+342420&   25.63 $\pm$     0.14&   24.55 $\pm$     0.31&   24.03 $\pm$     0.19&   14.98 $\pm$     2.33&   30.54 $\pm$     3.70&   80.92 $\pm$    14.62&  164.53 $\pm$    13.04\\
SST\_J142544.5+344558&   21.60 $\pm$     0.01&   25.59 $\pm$    0.10&   20.17 $\pm$     0.01&  129.37 $\pm$     3.88&  112.71 $\pm$     3.80&  132.03 $\pm$    17.19&  168.90 $\pm$    15.20\\
SST\_J142554.9+341820&   25.80 $\pm$     0.15&   24.87 $\pm$     0.23&   24.18 $\pm$     0.14&    9.39 $\pm$     1.92&   13.65 $\pm$     2.58&   11.16 $\pm$     7.86&   51.18 $\pm$     9.19\\
SST\_J142607.8+330425&   26.28 $\pm$     0.19&   24.61 $\pm$     0.12&   24.00 $\pm$     0.16&   32.04 $\pm$     3.34&   44.26 $\pm$     4.47&   75.84 $\pm$    13.80&  131.12 $\pm$    12.32\\
SST\_J142622.0+345249&   24.84 $\pm$     0.08&   24.46 $\pm$     0.16&   24.04 $\pm$     0.18&    4.29 $\pm$     1.41&    4.09 $\pm$     1.74&(10.0)&   36.97 $\pm$     7.70\\
SST\_J142626.4+344731&(27.1)&(26.1)&(25.4)&   18.26 $\pm$     2.69&   25.24 $\pm$     3.43&   39.80 $\pm$    12.05&   39.33 $\pm$     8.32\\
SST\_J142637.3+333025&(27.1)&(26.1)&   24.97 $\pm$     0.60&    4.42 $\pm$     1.45&   11.93 $\pm$     2.51&   34.77 $\pm$    11.32&   89.06 $\pm$    11.06\\
\enddata
\end{deluxetable}

\clearpage
\begin{deluxetable}{lccccllcccc}
\tabletypesize{\scriptsize}
\setlength{\tabcolsep}{0.05in} 
\tablecaption{Multi-Wavelength MIR and FIR Photometry, and Derived Properties (with limits in parentheses) \label{tab:photometry}}
\tablehead{\colhead{Galaxy ID}  &\colhead{24} &\colhead{250} &\colhead{350} &\colhead{500}  & \colhead{near-IR} & \colhead{mid/far-IR} & \colhead{\lir\tablenotemark{a}}  &\colhead{\lir\tablenotemark{a}} &\colhead{$IR8$\tablenotemark{b}} & \colhead{Temp}\\
&\colhead{[mJy]}&\colhead{[mJy]}&\colhead{[mJy]}&\colhead{[mJy]}& \colhead{Class} &\colhead{Class} &\colhead{SED} &\colhead{model\tablenotemark{c}} &&\colhead{[K]}}
\startdata
SST\_J142538.2+351855&    0.85 $\pm$     0.05&    56.0 $\pm$      6.1&    48.3 $\pm$      6.1&    35.3 $\pm$     10.0&      Bump&  NGC~6240&      1.21e+13&    1.08e+13&     10.89&     41\\
SST\_J142541.3+342420&    0.67 $\pm$     0.04&    29.3 $\pm$      3.0&    33.6 $\pm$      4.9&    26.6 $\pm$      7.6&        PL&  NGC~6240&      7.31e+12&    7.57e+12&     11.52&     32\\
SST\_J142544.5+344558&    0.77 $\pm$     0.04&    22.3 $\pm$      5.4& (20)& (25)&      Bump&  NGC~6240&      4.53e+11&    6.18e+11&     10.63&      -\\
SST\_J142554.9+341820&    1.14 $\pm$     0.05&    24.5 $\pm$      6.5&    36.1 $\pm$      5.8&    37.2 $\pm$      7.7&        PL&   Mrk~231&      4.88e+13&    6.97e+13&      4.87&     47\\
SST\_J142607.8+330425&    0.54 $\pm$     0.05& (20)& (20)& (25)&        PL&        - &    (3.05e+12)&    5.47e+12&    (9.78)&      -\\
SST\_J142622.0+345249&    1.30 $\pm$     0.05& (20)& (20)& (25)&        PL&        - &    (7.03e+12)&    1.18e+13&    (6.93)&      -\\
SST\_J142626.4+344731&    1.17 $\pm$     0.04&    55.3 $\pm$      5.3&    44.0 $\pm$      5.1&    33.9 $\pm$     10.6&      Bump&  NGC~6240&      1.11e+13&    1.20e+13&     17.20&     41\\
SST\_J142637.3+333025&    0.64 $\pm$     0.05& (20)& (20)& (25)&        PL&        - &    (1.36e+13)&    1.81e+13&    (6.55)&      -\\
\enddata
\tablenotetext{a}{\lsun}
\tablenotetext{b}{$IR8=L_{IR}/L_8$}
\tablenotetext{c}{\citet{Chary01}}
\end{deluxetable}

\end{document}